\documentclass[10pt,twocolumn,letterpaper]{article}

\usepackage{iccv}
\usepackage{times}
\usepackage{epsfig}
\usepackage{graphicx}
\usepackage{amsmath}
\usepackage{amssymb}

\usepackage{bm}
\usepackage{booktabs}
\usepackage{subcaption}
\usepackage{color}
\usepackage{dcolumn}
\usepackage{authblk}
\usepackage{appendix}
\newcommand{\tabincell}[2]{\begin{tabular}{@{}#1@{}}#2\end{tabular}}
\usepackage[breaklinks=true,colorlinks,bookmarks=false]{hyperref}

\iccvfinalcopy 

\def\iccvPaperID{****} 

\ificcvfinal\fi

\begin{document}

\title{Omniscient Video Super-Resolution}

\author[1]{Peng Yi}
\author[1]{Zhongyuan Wang\thanks{Corresponding author.}}%
\author[1]{Kui Jiang}
\author[2]{Junjun Jiang}
\author[3]{Tao Lu}
\author[1]{Xin Tian}
\author[1]{Jiayi Ma}
\affil[1]{Wuhan University $^2$Harbin Institute of Technology $^3$Wuhan Institute of Technology \authorcr {\tt\small \{yipeng, kuijiang, xin.tian\}@whu.edu.cn, wzy\_hope@163.com, jiangjunjun@hit.edu.cn, \{jyma2010, lutxyl\}@gmail.com}}

\maketitle
\ificcvfinal\thispagestyle{empty}\fi

\begin{abstract}
  Most recent video super-resolution (SR) methods either adopt an iterative manner to deal with low-resolution (LR) frames from a temporally sliding 
window, or leverage the previously estimated SR output to help reconstruct the current frame recurrently.
  A few studies try to combine these two structures to form a hybrid framework but have failed to give full play to it.
  In this paper, we propose an omniscient framework to not only utilize the preceding SR output, but also leverage the 
SR outputs from the present and future.
  The omniscient framework is more generic because the iterative, recurrent and hybrid frameworks can be regarded as its special cases.
  The proposed omniscient framework enables a generator to behave better than its counterparts under other frameworks.
  Abundant experiments on public datasets show that our method is superior to the state-of-the-art methods in objective metrics, subjective visual 
effects and complexity.
  Our code will be made public.
\end{abstract}

\section{Introduction} \label{sec:intro}
  Super-Resolution (SR) aims at reconstructing high-resolution (HR) images from the corresponding low-resolution (LR) images.
  As the most basic problem in SR, single image super-resolution (SISR) has been relatively studied thoroughly, where under a unified framework, 
the researchers only have to design different kinds of convolutional neural networks (CNNs) \cite{SRCNN, VDSR, zhangRDN, RCAN, daiSAN} to 
solve this issue.
  \begin{figure}[ht]
    \centering
      \includegraphics[width=0.35\textwidth]{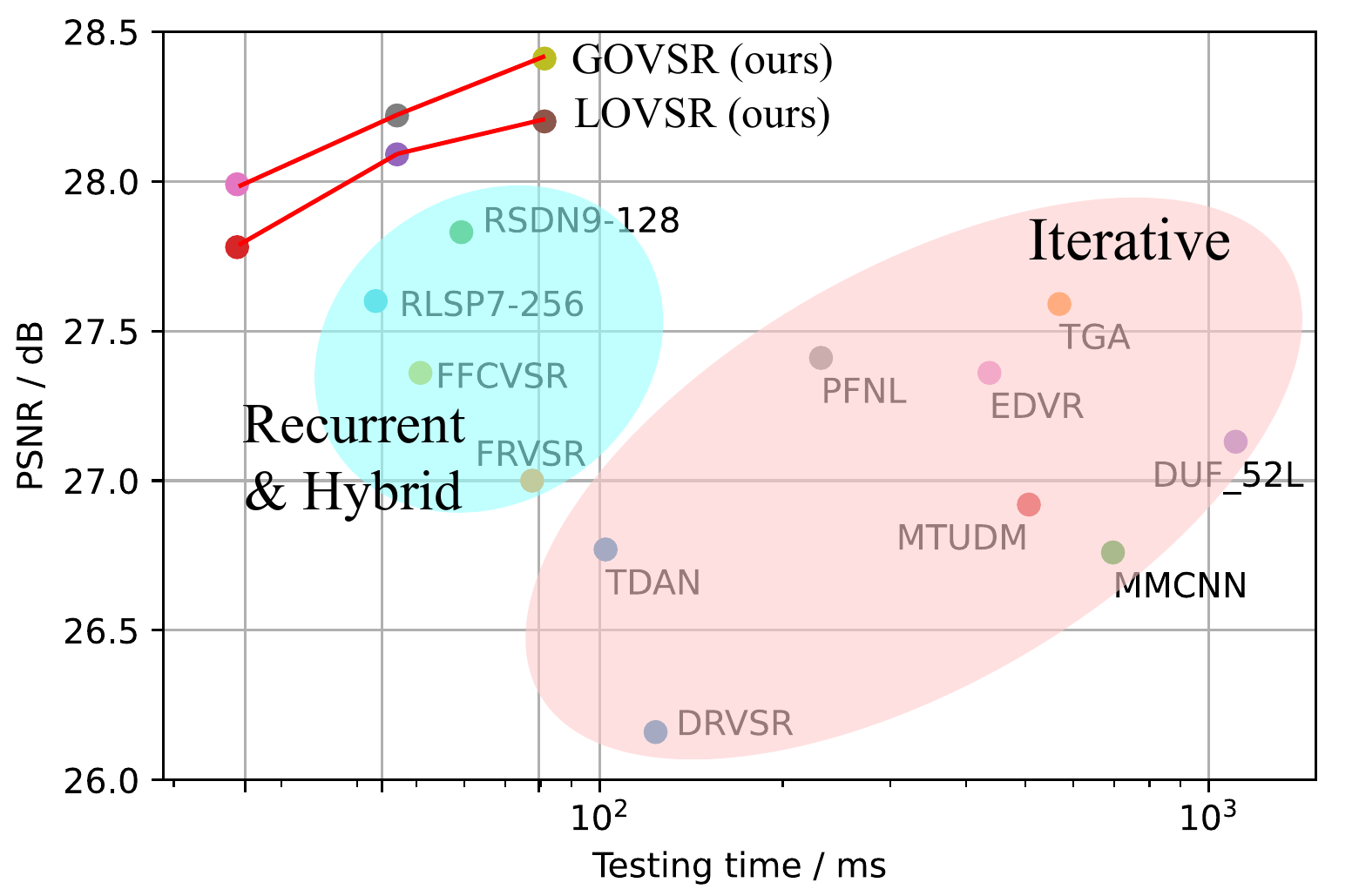}
    \caption{Performances on Vid4 \cite{VESPCN} dataset and time costs. The Red circle denotes iterative methods, and the blue 
    circle indicates recurrent and hybrid methods. Please refer to Table \ref{table:vid4} and Table \ref{table:comparison_sota} for more details.}
    \label{fig:testtime}
  \end{figure}
  Based on SISR, video super-resolution (VSR) has also been developed, albeit a lot of works \cite{VESPCN, DRVSR, FRVSR, FFCVSR, 
PFNL, TGA, RSDN} have been proposed, \emph{there is not a unified framework being dominant in VSR yet}.
  Figure \ref{fig:testtime} illustrates dozens of state-of-the-art (SOTA) VSR methods in terms of performance and speed.

\begin{figure*}[ht]
      \centering
      \begin{subfigure}[ht]{0.25\textwidth}
         \centering
         \includegraphics[width=1\textwidth]{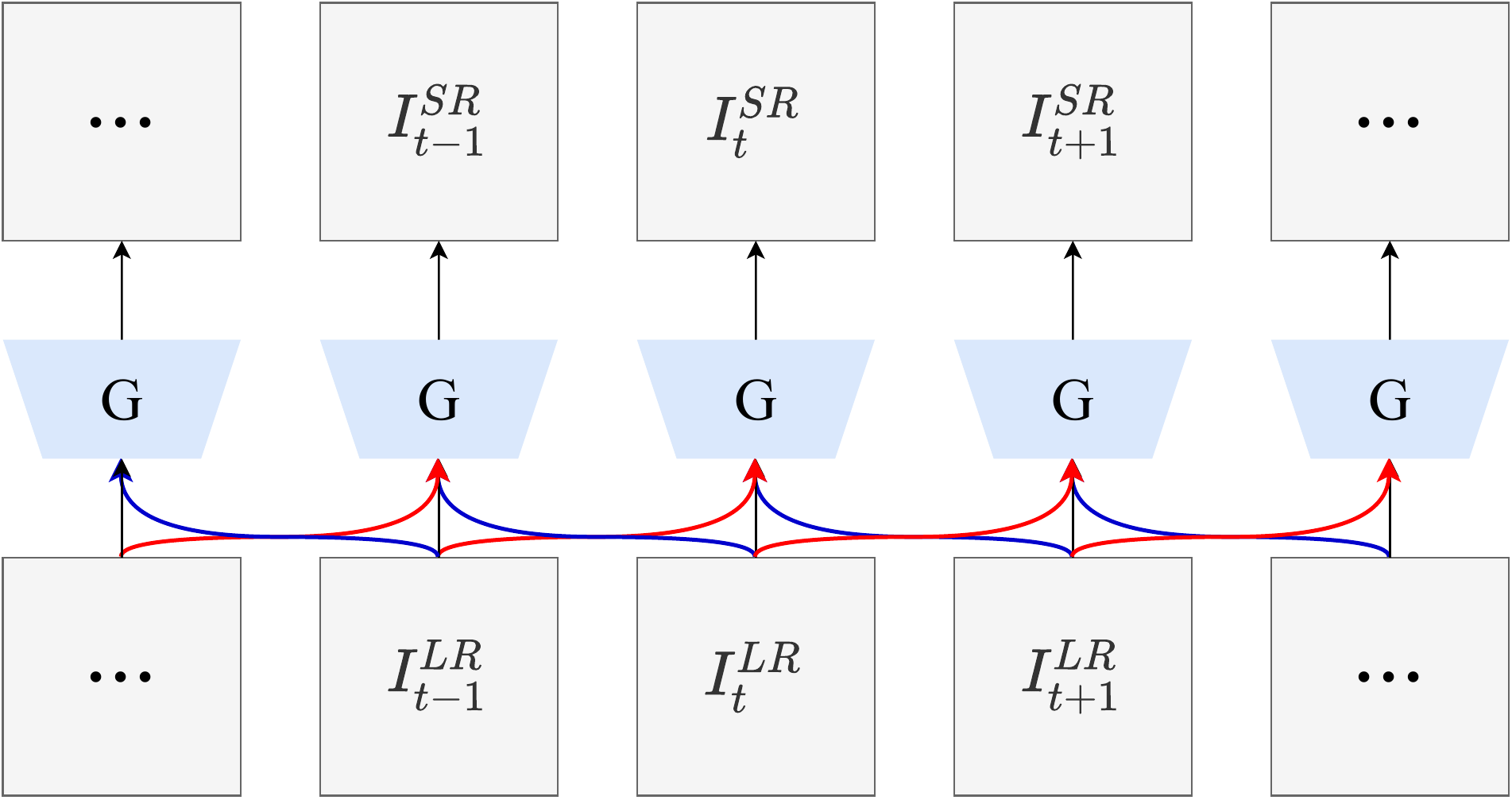}
         \caption{Iterative VSR}\label{fig:framework_ivsr}
      \end{subfigure}
      \quad
      \begin{subfigure}[ht]{0.25\textwidth}
         \centering
         \includegraphics[width=1\textwidth]{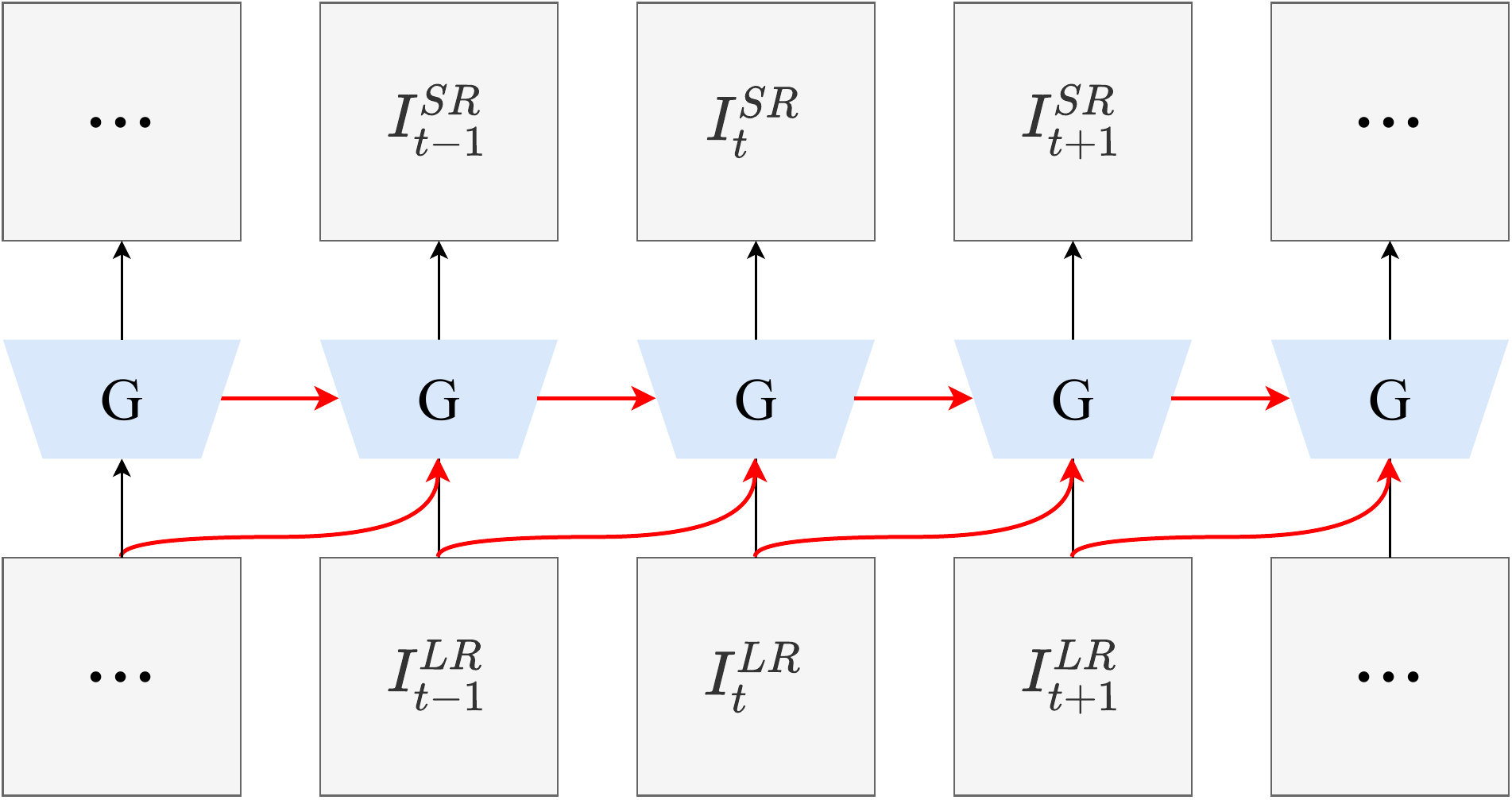}
         \caption{Recurrent VSR}\label{fig:framework_rvsr}
      \end{subfigure}
      \quad
      \begin{subfigure}[ht]{0.25\textwidth}
         \centering
         \includegraphics[width=1\textwidth]{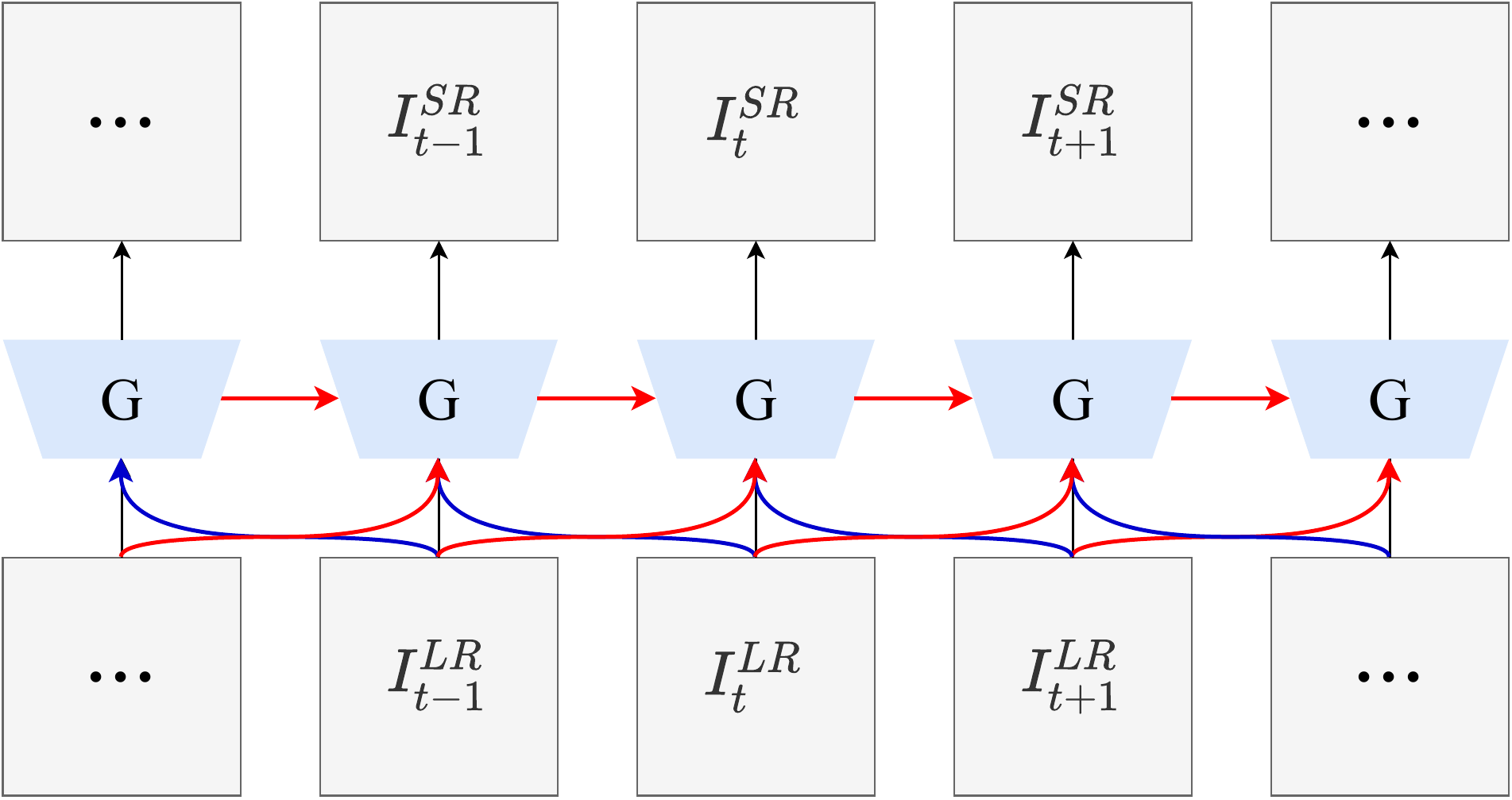}
         \caption{Hybrid VSR}\label{fig:framework_hvsr}
      \end{subfigure}

      \begin{subfigure}[ht]{0.35\textwidth}
         \centering
         \includegraphics[width=1\textwidth]{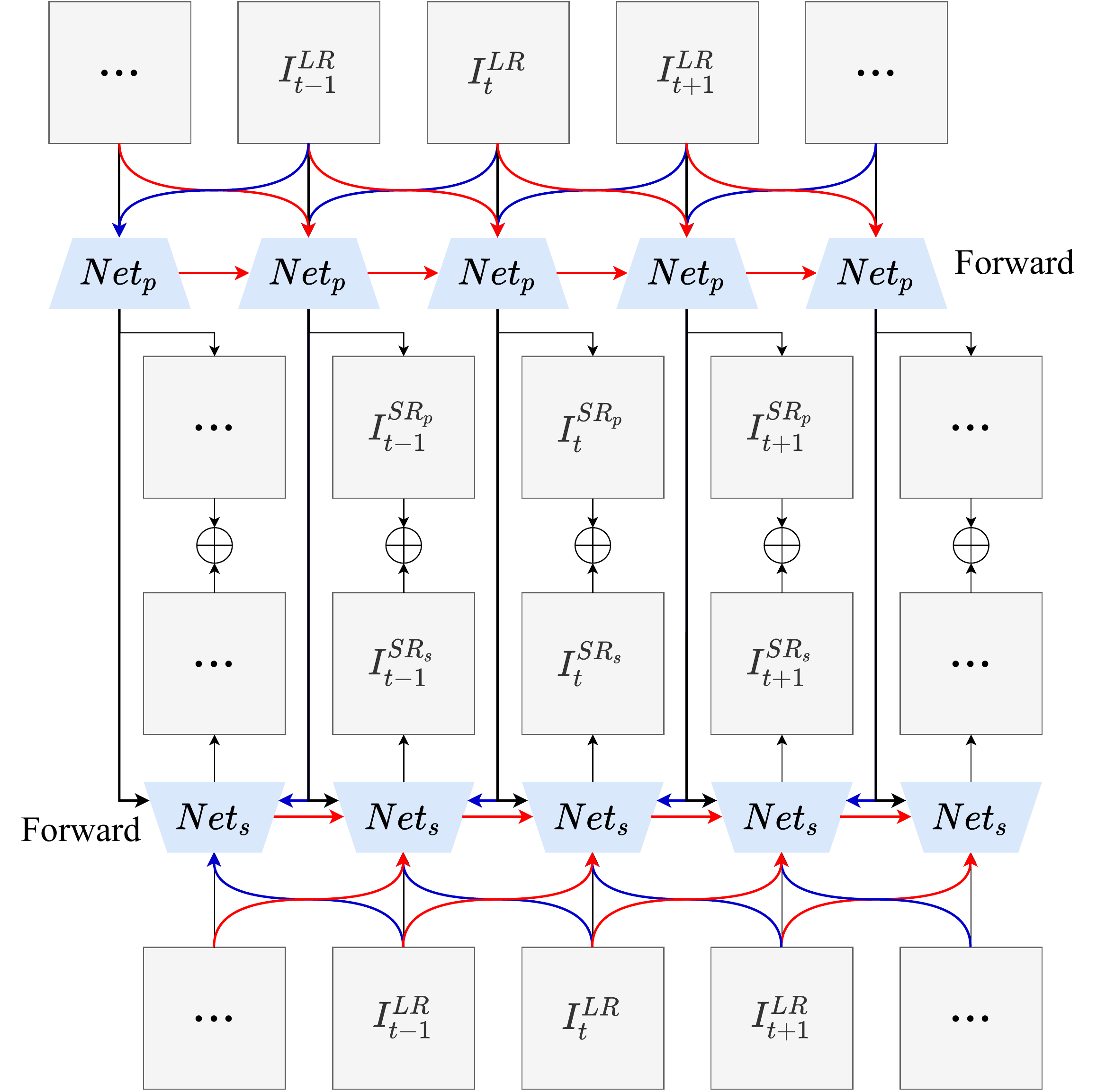}
         \caption{Local Omniscient VSR}\label{fig:framework_lovsr}
      \end{subfigure}
      \quad
      \quad
      \quad
      \begin{subfigure}[ht]{0.35\textwidth}
        \centering
        \includegraphics[width=1\textwidth]{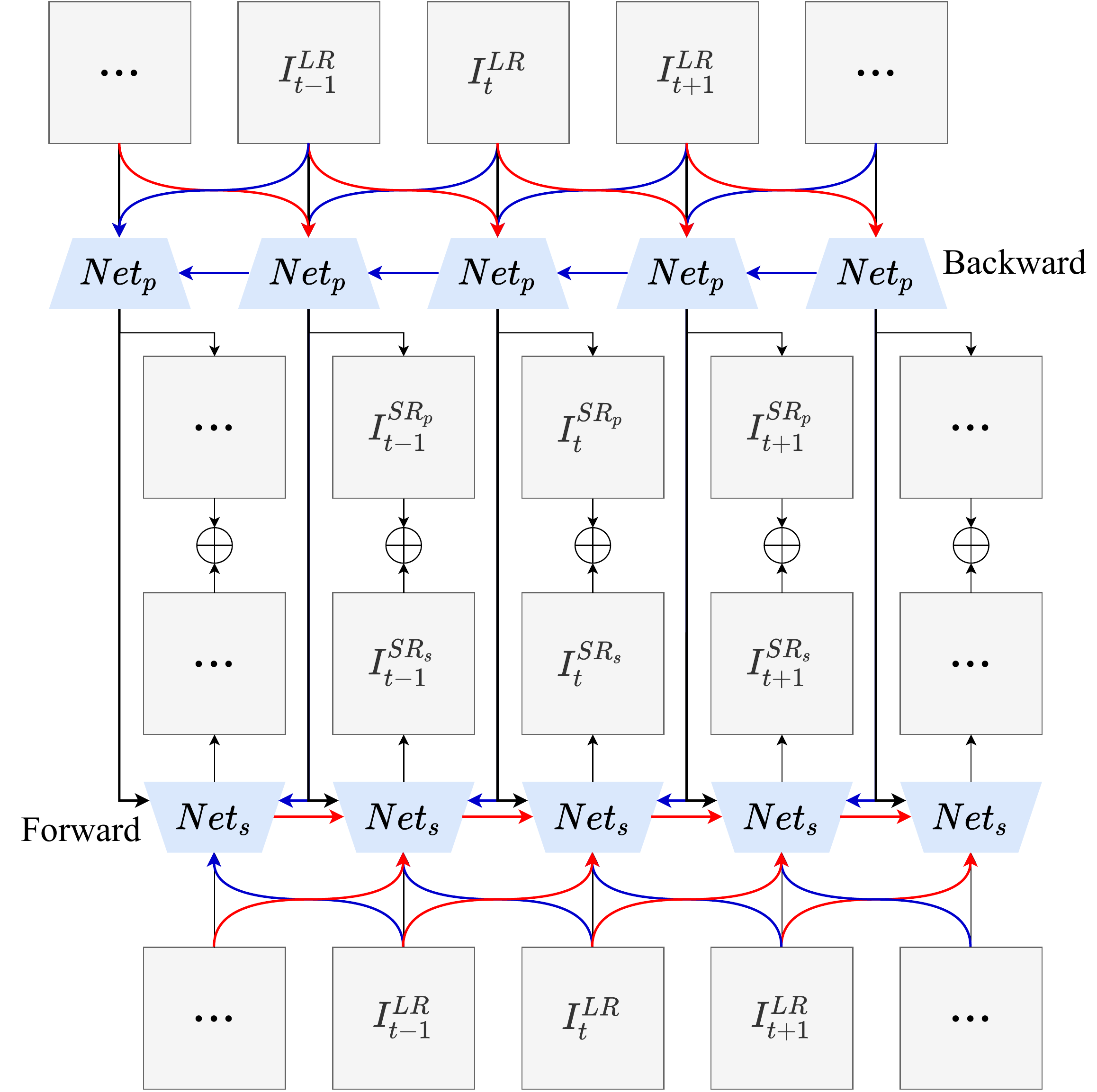}
        \caption{Global Omniscient VSR}\label{fig:framework_govsr}
      \end{subfigure}

      \caption{Different kinds of frameworks for VSR, where ``G'' represents the generator network. \textcolor{red}{Red}, black and \textcolor{blue}{blue} 
      arrows denote information from the past, present and future respectively.}
      \label{fig:framework_sr}
   \end{figure*}

  As SISR requires only one input image, most SISR methods focus on exploring different generator networks to extract features from this one image under a unified framework.
  Nevertheless, since VSR involves consecutive video frames as input, different schemes for handling the temporal information have emerged.
  We demonstrate different kinds of frameworks for VSR in Figure \ref{fig:framework_sr}.

  As illustrated in Figure \ref{fig:framework_ivsr}, most recent VSR methods \cite{VESPCN, DRVSR, MMCNN, MTUDM, DUFVSR, EDVR, PFNL, TDAN, TGA} apply 
an iterative manner to deal with LR frames from a temporally sliding window, where we only show the case of window size as 3.
  Given a sequence of video frames, the iterative framework considers the whole VSR processing as multiple independent sub-processes.
  Theoretically, these sub-processes are not temporally correlated and can be handled simultaneously, which means they enjoy the advantage of parallel 
computing \cite{PFNL}.
  However, the iterative framework can only obtain more neighboring LR frames by increasing the window size but omits the previously estimated SR 
output, which is the exact reason that prevents it from a better performance.

  As demonstrated in Figure \ref{fig:framework_rvsr}, the recurrent framework \cite{FRVSR, RSDN} processes video frames by the order, while it can 
never exploit the subsequent frames to assist with recovering the current frame, which has limited its potential.
  Although a few studies \cite{RLSP, FFCVSR} have tried to combine these two frameworks to form a hybrid framework, as shown in Figure 
\ref{fig:framework_hvsr}, the hybrid framework can only receive the estimated hidden states from the past, and they have not achieved satisfying results.

  The recurrent and hybrid frameworks only leverage previous hidden states, which inspired us to think about \emph{what if we further try to involve the hidden states from the present and future}.
  Thus, after careful consideration, we propose the omniscient framework.
  Specifically, we integrate two sub-networks: a precursor network $Net_p$ and a successor network $Net_s$ into the omniscient framework.
  The successor network inherits the hidden states generated by the precursor network, and thus manages to leverage the \emph{LR frames and hidden states from the past, present and future}.
  As shown in Figure \ref{fig:framework_lovsr} and Figure \ref{fig:framework_govsr}, the omniscient framework can be further divided into two categories: local omniscient and global omniscient.
  Local omniscient framework processes video frames unidirectionally while global omniscient framework does it bidirectionally.
  The global omniscient framework enables any LR frame to receive information from all other frames in a same video sequence, however, it is not appropriate for delay-sensitive real-time tasks 
like live broadcasts, where the local omniscient framework suits well instead.
  
  Overall, in this paper, we propose a more generic omniscient framework to exploit both LR frames and estimated hidden states from the past, present and future.
  In fact, as shown in Figure \ref{fig:framework_sr}, \emph{the iterative, recurrent and hybrid frameworks can be regarded as the special counterparts of our 
proposed omniscient framework.}
  We have explored a same kind of generator network under different frameworks, and we have found the omniscient framework (local and global) superior to the existing iterative, recurrent 
and hybrid frameworks.
  Our models surpass other SOTA methods in both performance and complexity, and thus \emph{we hope this framework will become a standard framework in VSR, under which researchers are free 
to design more effective generator networks, explicit or implicit motion information capturing modules, or loss functions to tap its potential.}

\section{Related Work} \label{sec:related_work}

\subsection{Iterative Video Super-Resolution} \label{sec:rw_ivsr}
  With the development of deep learning, a lot of CNN-based SISR methods, \eg SRCNN \cite{SRCNN}, VDSR \cite{VDSR}, have emerged, from which some early VSR 
methods \cite{VESPCN, VSRnet, MCResNet} are inspired and simply apply the architectures from SISR to VSR.
  These methods consider VSR as a multi-input counterpart of SISR and employ the iterative framework to solve it.
  These iterative methods can be further divided into two categories according to their generator networks: iterative \cite{RVSR-LTD, DUFVSR, EDVR, PFNL, TDAN, TGA, PFNL-GAN} 
and recurrent \cite{DRVSR, MMCNN, MTUDM, RBPN}.
  Particularly, RVSR-LTD \cite{RVSR-LTD} designs a temporal adaptive network with a pyramid structure, and DUFVSR \cite{DUFVSR} proposes a 3D CNN with dynamic 
upsampling filters.
  EDVR \cite{EDVR} has designed a pyramid, cascading and deformable (PCD) alignment module with a temporal and spatial attention (TSA) fusion module to 
achieve wonderful results, however, its training resource requirement is prohibitive.
  PFNL \cite{PFNL} proposes a progressive fusion network with a non-local correlations extraction module, which has obtained an amazing result in terms of 
both performance and complexity.
  TDAN \cite{TDAN} devises a temporally-deformable alignment network to avoid the optical flow estimation.
  TGA \cite{TGA} divides temporal information into groups, and utilizes a hybrid module with both 2D and 3D residual blocks for inter-group fusion.
  
  Another group of iterative-based methods adopts recurrent generator networks to transfer temporal correlations between frames instead.
  DRVSR \cite{DRVSR} utilizes ConvLSTM \cite{ConvLSTM} module for capturing long-range temporal information, based on which MMCNN \cite{MMCNN} designs a 
multi-memory residual block to enhance the memory ability.
  MTUDM \cite{MTUDM} proposes an ultra-dense memory residual block combined with a multi-temporal information fusion strategy to build a shallower but wider network.
  RBPN \cite{RBPN} extends \cite{DBPN} to video SR, which sends LR frames into a projection module step by step according to their temporal distances.
  \emph{In all, these iterative methods have been concentrating on designing more and more complicated generator networks but with huge computational costs in the meantime.}
  
\subsection{Recurrent and Hybrid Video Super-Resolution} \label{sec:rw_rhvsr}
  As to the recurrent methods, FRVSR \cite{FRVSR} proposes a frame-recurrent network to leverage the last rebuilt SR frame for reconstructing the current 
frame, which is fast but not robust enough.
  RSDN \cite{RSDN} designs a dual-channel network to learn the structures and details of frames.
  \emph{Still, these recurrent methods refuse the assist from subsequent LR frames.}

  FFCVSR \cite{FFCVSR} and RLSP \cite{RLSP} have tried to form a hybrid framework but could not achieve a wonderful result because 1) the hybrid framework 
still cannot exploit the estimated hidden states from the present and future, and 2) their naive network designs have turned them against maximizing their 
potential.

\begin{figure*}[ht]
   \centering
     \includegraphics[width=0.99\textwidth]{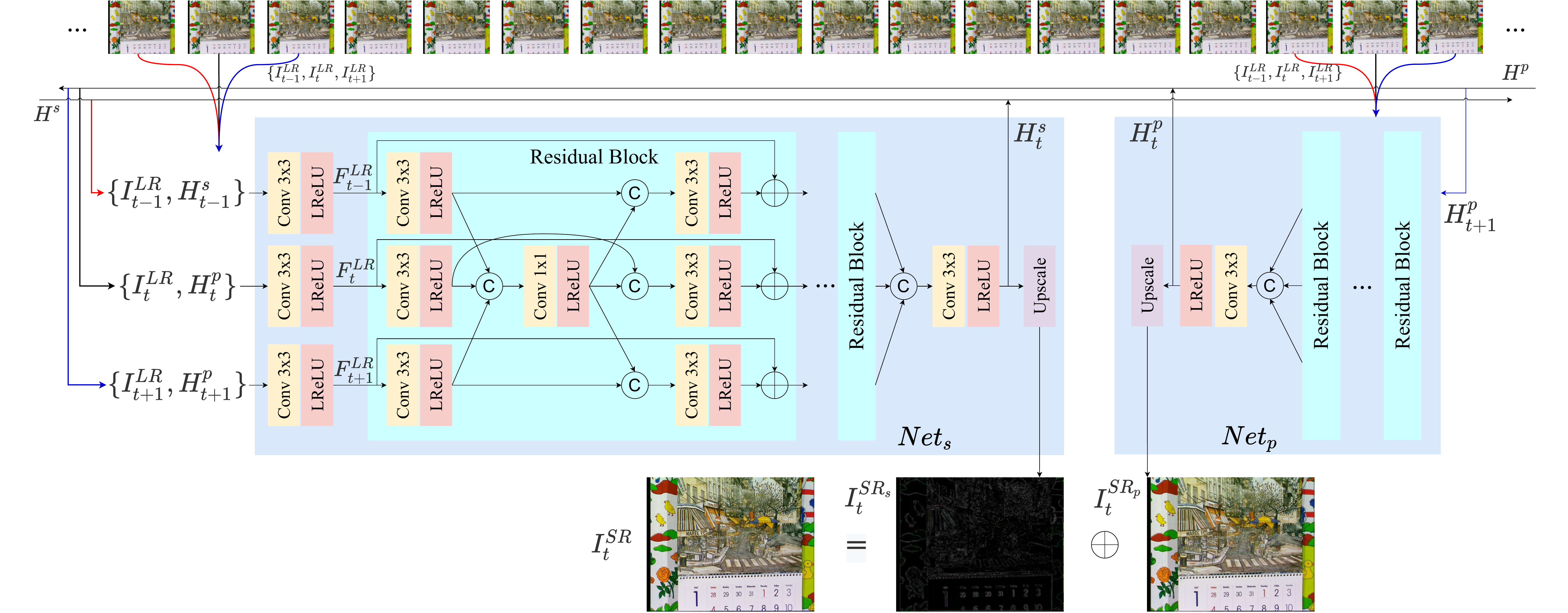}
   \caption{The whole architecture of our model under the global omniscient framework, and the local omniscient case can be inferred according to Figure \ref{fig:framework_lovsr}. 
   The ``$\bigoplus$'' represents element-wise add and ``$\copyright$'' denotes concatenation along the 
   channel axis. \textcolor{red}{Red}, black and \textcolor{blue}{blue} arrows denote information from the past, present and future respectively.}
   \label{fig:pfovsr}
\end{figure*}

\section{Method} \label{sec:method}

\subsection{Omniscient Video Super-Resolution} \label{sec:method_ovsr}
  We first introduce the core idea of omniscient video super-resolution (OVSR).
  As illustrated in Figure \ref{fig:framework_sr}, the hybrid VSR combines the idea of iterative VSR and recurrent VSR, which adopts both neighboring LR frames and 
a previously estimated SR output as source information.
  Theoretically, the neighboring LR frames provide \emph{the most basic spatial-temporal information in LR space}, and the estimated SR output can reserve more 
\emph{temporally correlated information connected to the HR space}.
  Thus, it is natural to combine them to fully exploit the spatial-temporal correlations.

  Based on the above analysis, we wonder what if we further introduce the estimated SR outputs from the present and future.
  However, after long deliberation, we reckon that idea is not realizable if only processing through the video frames in a single time like the iterative, recurrent, or hybrid VSR.
  Eventually, as shown in Figure \ref{fig:framework_lovsr} and Figure \ref{fig:framework_govsr}, we manage to design two sub-networks: a precursor 
network $Net_p$ and a successor network $Net_s$ to form the OVSR framework.
  The precursor network first flows through the LR frames to generate the SR frames and hidden states of all time steps, and then the successor network reconstructs 
all SR frames with the assist of the corresponding LR frames and estimated hidden states.
  We further add the SR frames generated both by the precursor and successor for refinement to reconstruct the final SR output.

  According to the direction of the precursor network and successor network, the OVSR can be further categorized as local omniscient VSR (LOVSR) 
and global omniscient VSR (GOVSR).
  Both $Net_p$ and $Net_s$ of LOVSR process video frames in the same direction, which means it can only utilize the past and present information, as well as a limited 
number of frames (2 in Figure \ref{fig:framework_lovsr}) in the future, which is why we call it ``local omniscient''.
  Then, we design the GOVSR by reversing the direction of $Net_p$ subtly, through which any LR frame gets access to all other frames in a same video sequence, and thus 
it is called ``global omniscient''.
  Although the GOVSR is capable of leveraging all frames to assist in rebuilding one frame at any time step, it relies heavily on future information.
  Nevertheless, frames too far in the future are actually not accessible in some delay-sensitive real-time tasks like live broadcasts and online meetings, where 
the LOVSR is more appropriate.
  In all, \emph{LOVSR suits online VSR and GOVSR is more appropriate for offline VSR.}
  Still, the omniscient VSR, no matter LOVSR or GOVSR, can exploit both the LR frames and estimated hidden states from the past, present and future.

  It is worth mentioning that the proposed omniscient framework \emph{does not require a specific structure of the generator network, instead, any kind of 
existing generator networks can be easily inserted into the architecture illustrated in Figure \ref{fig:pfovsr}, as long as they satisfy the form shown in 
Equation \eqref{eq:precursor_net} and Equation \eqref{eq:successor_net}.}
  
\subsection{Network Design} \label{sec:method_netdesign}
  We then elaborate on the specific network design of our model under the omniscient framework.
  Although our proposed OVSR manages to leverage more source information theoretically, it still requires a well-designed structure of the generator 
network to fulfill its potential practically.
  Fortunately, we learn a progressive fusion residual block (PFRB) from PFNL \cite{PFNL}, which is sophisticated and shows to be both effective and efficient.

  To combine the PFRB with the omniscient framework, we ameliorate it to embody 3 channels.
  As illustrated in Figure \ref{fig:pfovsr}, we show the architecture of our model under GOVSR, and the LOVSR case can be inferred.
  Because the precursor and successor networks share a similar structure, we only need to introduce the successor network.
  We adjust the PFRB to embody 3 channels, which contain information from the past, present and future respectively.
  We first adopt one convolutional layer to fuse the corresponding LR frame $I_t^{LR}$ and hidden state $H_t$ to obtain a feature $F_t^{LR}$.
  Correspondingly, these 3 features contain information from the past, present and future respectively.
  Then, in the residual blocks, features from these 3 channels are extracted both independently and merged together, in which intra-frame spatial 
correlations and inter-frame temporal correlations are fully exploited \cite{PFNL}.
  At the end of the network, features from these 3 channels are concatenated and processed by a $3\times 3 $ convolutional layer to obtain the updated 
hidden state $H_t^{s}$.
  The $H_t^{s}$ is upscaled to $I_t^{SR_s}$, and then added by the SR frame from the precursor $I_t^{SR_p}$ to reconstruct the final SR output $I_t^{SR}$, 
where the upscale module is composed of 2 convolutional layers each followed by a sub-pixel convolution layer \cite{ESPCN2016}.
  We set a Leaky ReLU activation \cite{He2016Delving} after every convolutional layer (except the last one in the upscale module), with 
parameter $\alpha=0.2$.

  The precursor network can be described as
  \begin{equation}
    I_t^{SR_p},H_t^{p}=Net_p(\{I_{t-1}^{LR},I_t^{LR},I_{t+1}^{LR}\},H_{t+1}^{p}), \label{eq:precursor_net}
  \end{equation}
  where $\{I_{t-1}^{LR},I_t^{LR},I_{t+1}^{LR}\}$ stand for the neighboring LR frames, $I_t^{SR_p}$ and $H_t^{p}$ denote the SR frame and hidden state generated 
by the precursor.
  Note that $H_{t+1}^{p}$ is for GOVSR, which should be $H_{t-1}^{p}$ for LOVSR instead.
  The successor network can be described as
  \begin{equation}
    I_t^{SR_s},H_t^{s}=Net_s(\{I_{t-1}^{LR},I_t^{LR},I_{t+1}^{LR}\},\{H_{t-1}^{s},H_{t}^{p},H_{t+1}^{p}\}), \label{eq:successor_net}
  \end{equation}
  where $I_t^{SR_s}$ and $H_t^{s}$ denote the SR frame and hidden state generated by the successor.
  We add the SR frames generated by $Net_p$ and $Net_s$ to form the final output.
  \begin{equation}
   I_t^{SR}=I_t^{SR_s}+I_t^{SR_p}, \label{eq:p+s}
  \end{equation}
  because $Net_s$ inherits $Net_p$, we naturally restrict $Net_p$ to mainly learn the low-frequency structures and $Net_s$ to study the high-frequency 
details within frames.
  We adopt the Charbonnier loss function \cite{LapSRN} to form the loss function:
  \begin{equation}
    \mathcal{L}=\sqrt{(I_t^{HR}-I_t^{SR})^2+\varepsilon^2} + \alpha \sqrt{(I_t^{HR}-I_t^{SR_p})^2+\varepsilon^2}, \label{eq:loss}
  \end{equation}
  where $I_t^{HR}$ denotes the original HR frame, $\alpha$ is set to adjust the weight of the precursor network, and $\varepsilon$ is empirically set to $10^{-3}$.

  In all, our model fully exploits the spatio-temporal correlations contained in the LR frames and estimated hidden states from the past, present and future.

\begin{figure*}[ht]
   \centering
   \begin{subfigure}[t]{0.3\textwidth}
      \centering
      \includegraphics[width=1\textwidth]{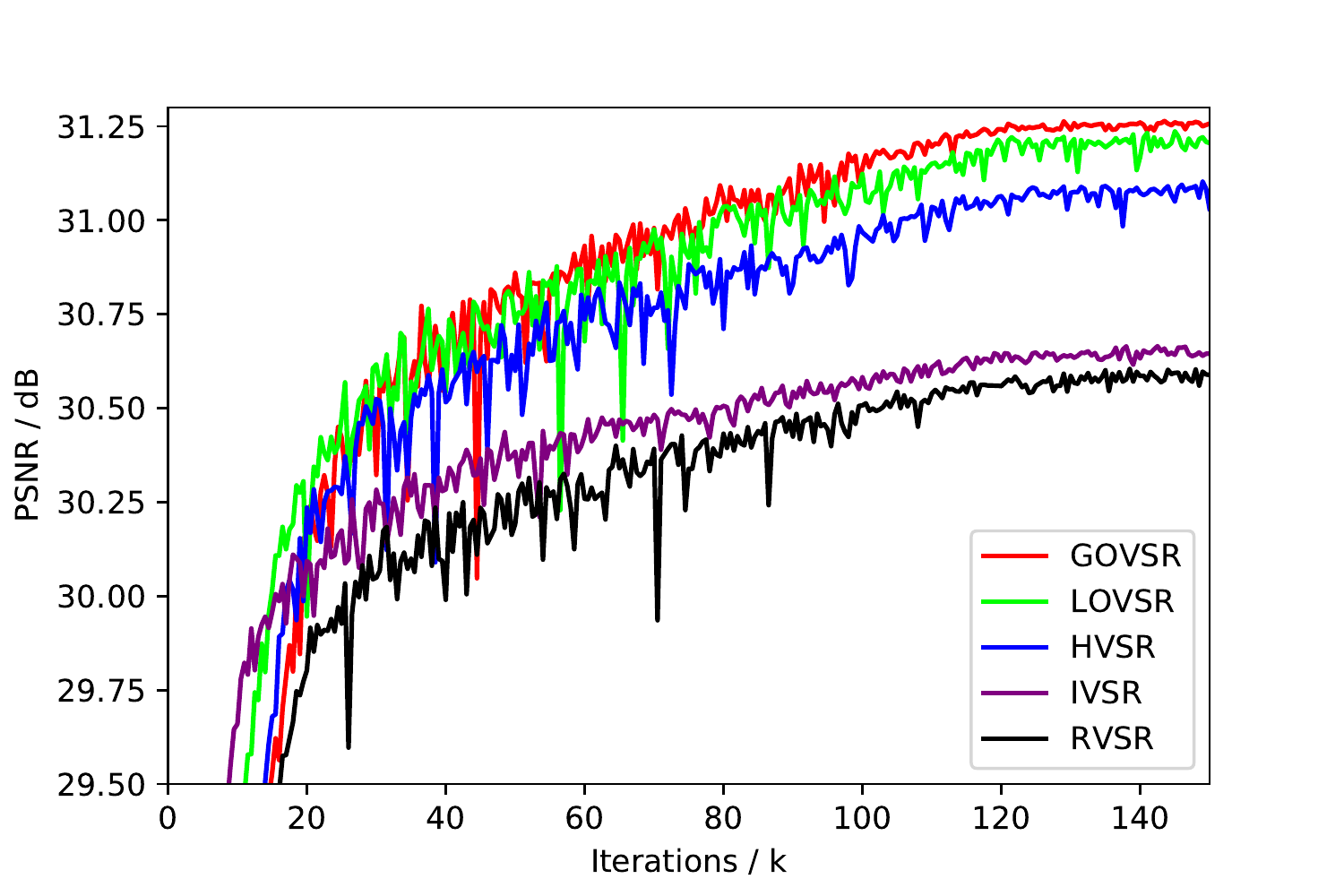}
      \caption{}\label{fig:abl_pfrb}
   \end{subfigure}
   \begin{subfigure}[t]{0.3\textwidth}
      \centering
      \includegraphics[width=1\textwidth]{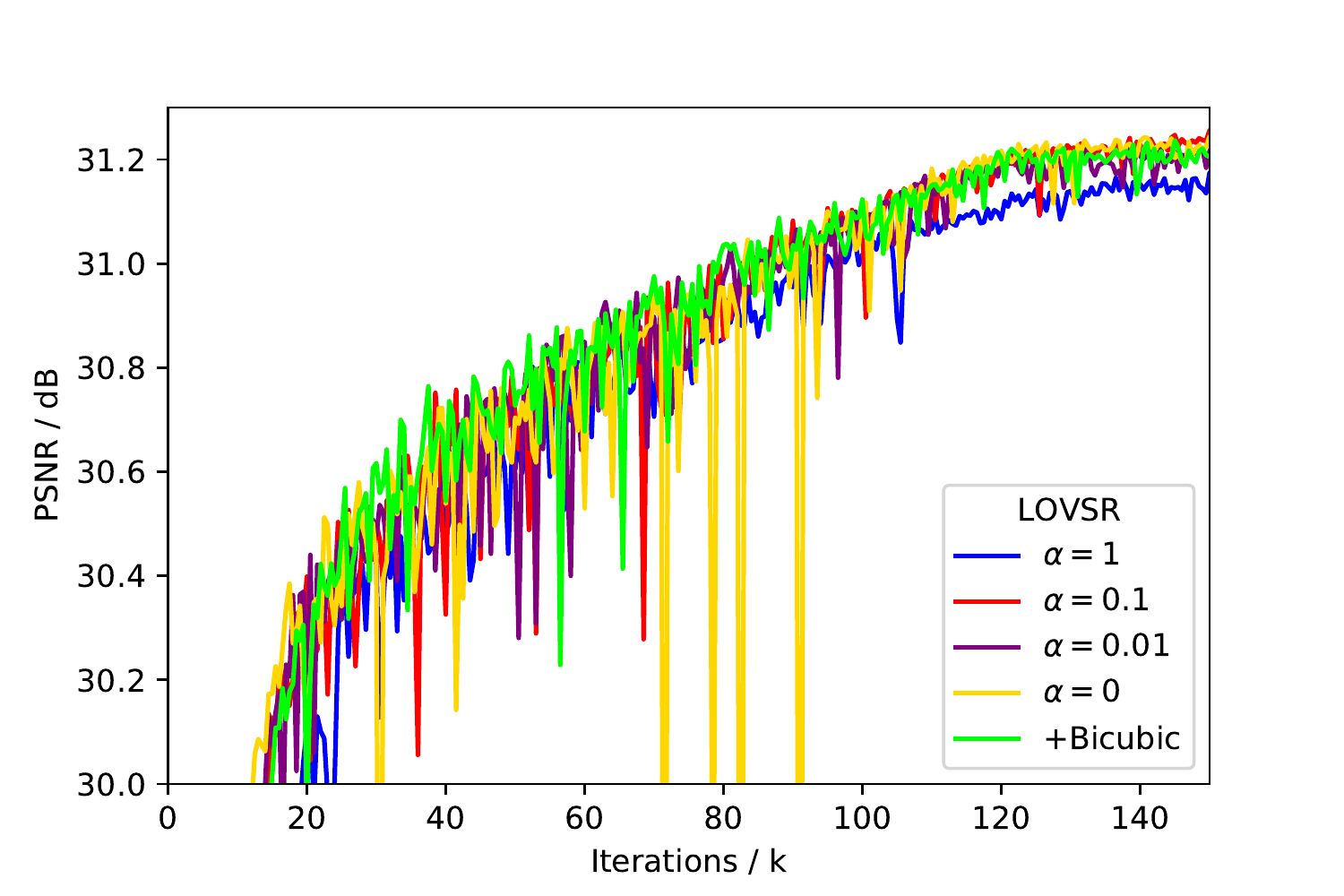}
      \caption{}\label{fig:abl_lovsr_p+s}
   \end{subfigure}
   \begin{subfigure}[t]{0.3\textwidth}
      \centering
      \includegraphics[width=1\textwidth]{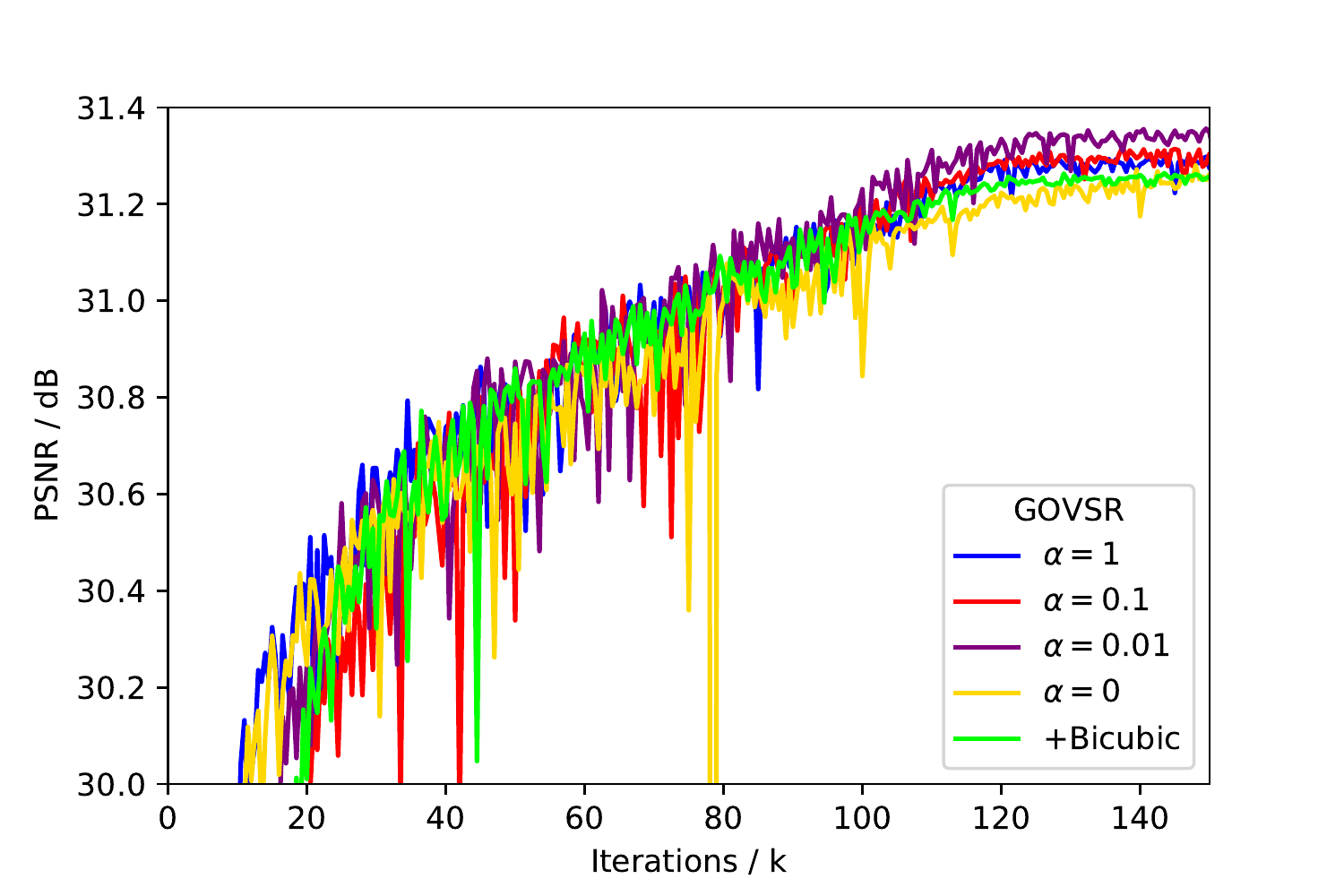}
      \caption{}\label{fig:abl_govsr_p+s}
   \end{subfigure}

   \begin{subfigure}[t]{0.3\textwidth}
      \centering
      \includegraphics[width=1\textwidth]{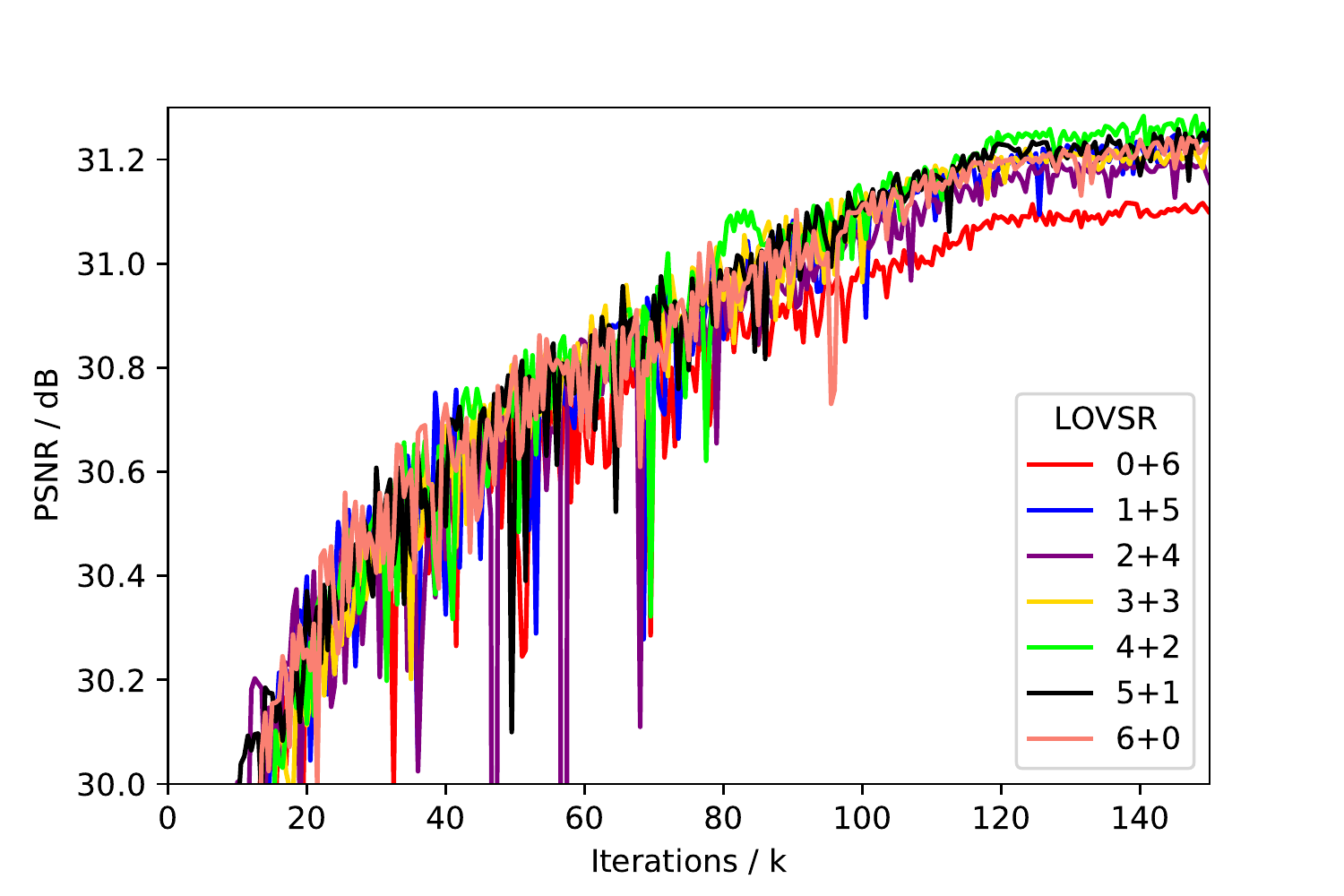}
      \caption{}\label{fig:abl_lovsr_proportion}
   \end{subfigure}
   \begin{subfigure}[t]{0.3\textwidth}
      \centering
      \includegraphics[width=1\textwidth]{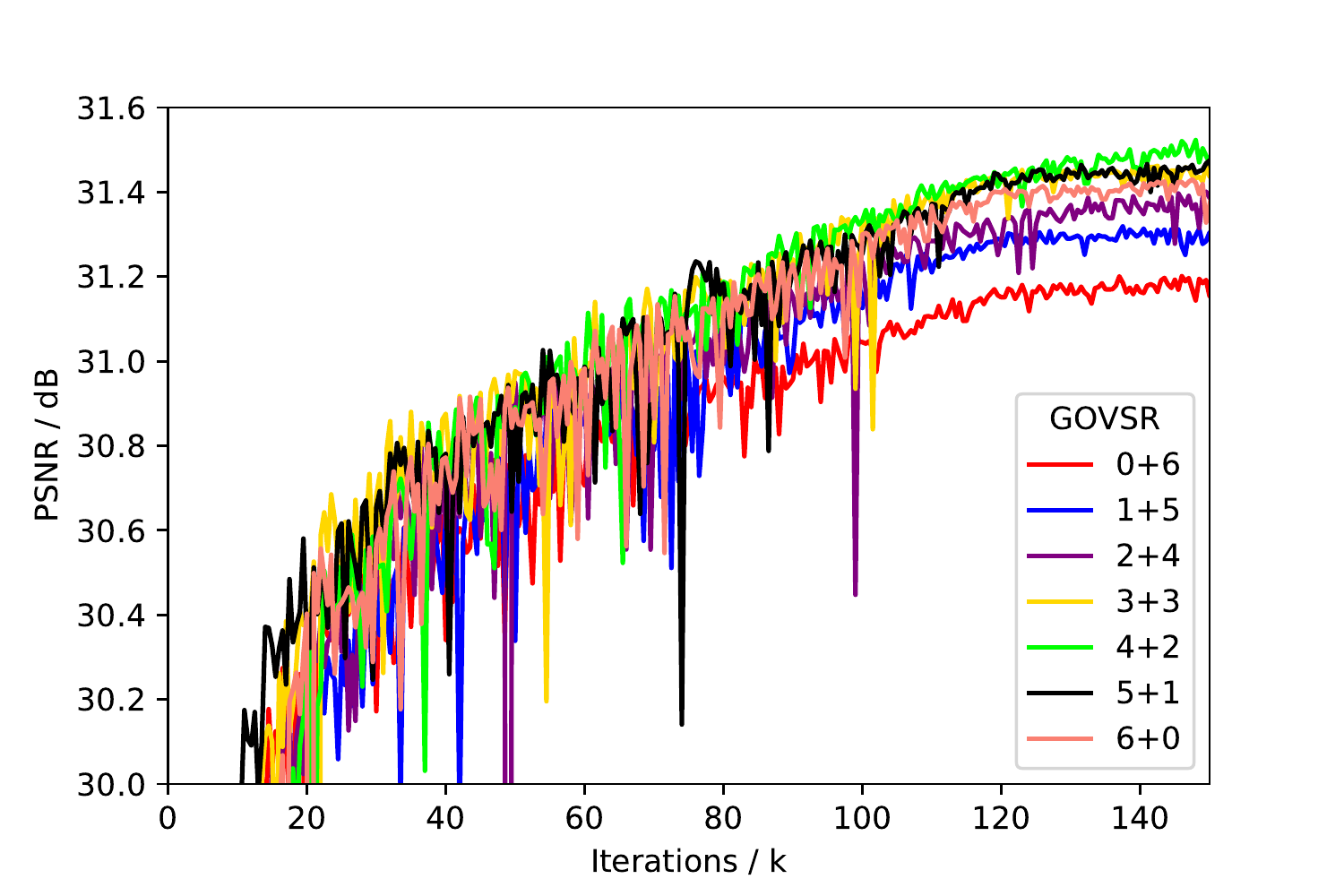}
      \caption{}\label{fig:abl_govsr_proportion}
   \end{subfigure}
   \caption{(a) Similar generator networks trained under different VSR frameworks. 
   (b)(c) Adjust the weight of $I_t^{SR_p}$ during training by changing $\alpha$ in Equation \eqref{eq:loss}. 
   (d)(e) Adjust the proportion of $Net_p$ and $Net_s$, where ``4+2'' means setting 4 residual blocks in $Net_p$ and 2 in $Net_s$.}
   \label{fig:abl_framework}
 \end{figure*}

\section{Experiments} \label{sec:experiments}
  
\subsection{Implementation Details} \label{sec:exp_id}
  We first adopt a public MM522 dataset \cite{MMCNN} for training, which contains 522 32-frame sequences with various scenes.
  During training, we employ 20 video sequences from \cite{PFNL} for evaluation.
  When trained on the MM522 dataset, we test the models on Vid4 \cite{VESPCN} and UDM10 \cite{MTUDM} testing datasets.
  We also conduct experiments on another public training dataset Vimeo-90K \cite{TOFLOW}, and test the models on its testing dataset Vimeo-90K-T (please see 
the appendix).
  Like in \cite{DUFVSR, EDVR, PFNL, RSDN}, we adopt Gaussian blur with $\sigma=1.6$ and then $4\times$ down-sampling scheme.

  We use Adam \cite{adam} optimizer with $\beta_1=0.9$ and $\beta_2=0.999$.
  We set the batch size as 16 and input LR size as $64\times 64$.
  We only feed 7 consecutive frames for the ablation study, and to train our final models as well as rebuild other SOTA methods, we set 2 additional 
frames for feeding the $I_{t-1}^{LR}, H_{t-1}$ in the beginning and $I_{t+1}^{LR}, H_{t+1}$ at the end \cite{RLSP}.
  The initial learning rate is $1\times 10^{-3}$ and decays linearly to $1\times 10^{-4}$ after 120 K iterations, which keeps the same until 200 K iterations.
  Then, the learning rate is further decayed to $5\times 10^{-5}$ and $1\times 10^{-5}$ until converged.
  The whole training process lasts about 300 K - 400 K iterations.
  We conduct experiments on Python 3.6, PyTorch 1.6 and NVIDIA RTX 2080Ti GPUs.

  \begin{table}[ht]
   \caption{Performances, parameters, FLOPs and testing time costs of different models. The PSNR is evaluated on 20 sequences from \cite{PFNL}, while FLOPs and 
   testing time costs are evaluated with respect to $1280\times 720$ HR frame under $4\times$ SR.}
   \renewcommand{\arraystretch}{1.2}
   \label{table:abl_framework}
   \centering
   \scriptsize
   \begin{tabular}{ccccccccc}
     \toprule
     Model             & IVSR      & RVSR      & HVSR      & LOVSR    & GOVSR    \\
     \midrule
     Parameter (M)     & 1.864     & 1.866     & 1.868     & 1.897    & 1.897    \\
     FLOPs (G)         & 107.780   & 107.696   & 107.796   & 109.746  & 109.746  \\
     Testing time (ms) & 23.06     & 23.33     & 23.29     & 25.35    & 25.35    \\
     PSNR (dB)         & 30.66     & 30.60     & 31.10     & 31.24    & 31.27    \\
     \bottomrule
   \end{tabular}
   \end{table}

\subsection{Omniscient vs Iterative, Recurrent and Hybrid} \label{sec:exp_frameworks}
  We compare the same generator network under the iterative, recurrent, hybrid and omniscient frameworks (denoted as IVSR, RVSR, HVSR and OVSR), where only 
necessary adjustments are made to satisfy the form of input/output of these frameworks.
  Note that IVSR, RVSR and HVSR only embody one generator network, and thus we have to add the Bicubic amplified frame $I_t^{Bic}$ for residual learning \cite{He2015Deep,VDSR}.
  For LOVSR and GOVSR, we also replace $I_t^{SR_p}$ with $I_t^{Bic}$ in Equation \eqref{eq:p+s}, and we set $\alpha=0$ in Equation \eqref{eq:loss} for a fair comparison.
  We integrate the multi-channel PFRBs into these frameworks, where we set 5 PFRBs with the filter number as 64 for IVSR, RVSR and HVSR.

  \begin{table}[ht]
    \caption{PSNR (dB) evaluated by adjusting the weight of $I_t^{SR_p}$ through changing $\alpha$ in Equation \eqref{eq:loss}.}
    \renewcommand{\arraystretch}{1.2}
    \label{table:abl_refine}
    \centering
    \scriptsize
    \begin{tabular}{ccccccccc}
      \toprule
      Setting           & +Bicubic  & $\alpha=0$ & $\alpha=0.01$ & $\alpha=0.1$ & $\alpha=1$ \\
      \midrule
      LOVSR             & 31.24     & 31.25      & 31.23         & 31.26        & 31.17      \\
      GOVSR             & 31.27     & 31.28      & 31.36         & 31.32        & 31.30      \\
      \bottomrule
    \end{tabular}
    \end{table}

    \begin{table}[ht]
       \caption{PSNR (dB) evaluated by adjusting the residual blocks in the precursor and successor networks. ``4+2'' denotes setting
       4 residual blocks in $Net_p$ and 2 in $Net_s$}
       \renewcommand{\arraystretch}{1.2}
       \label{table:abl_proportion}
       \centering
       \scriptsize
       \begin{tabular}{cccccccccc}
         \toprule
         Setting           & 0+6     & 1+5      & 2+4      & 3+3     & 4+2     & 5+1    & 6+0   \\
         \midrule
         LOVSR             & 31.12   & 31.26    & 31.20    & 31.23   & 31.28   & 31.26  & 31.24 \\
         GOVSR             & 31.20   & 31.36    & 31.40    & 31.46   & 31.52   & 31.47  & 31.43 \\
         \bottomrule
       \end{tabular}
       \end{table}

  As shown in Figure \ref{fig:abl_pfrb} and Table \ref{table:abl_framework}, HVSR outperforms IVSR and RVSR a lot in PSNR, which confirms the effectiveness of 
utilizing a hidden state from the past and a LR frame of the future.
  Nevertheless, as has been discussed in Section \ref{sec:intro} and Section \ref{sec:method_ovsr}, HVSR still fails to leverage the hidden states from the present and future.
  Then, to train our models LOVSR and GOVSR, we first set only 1 PFRB in $Net_p$ and 5 PFRBs in $Net_s$, and we adjust the filter number as 56.
  This setting aims to keep the parameters, calculation and time costs of these 5 models almost the same.
  As shown in Table \ref{table:abl_framework}, under similar parameters, LOVSR and GOVSR surpasses HVSR about 0.14 dB and 0.17 dB in PSNR respectively.
  GOVSR enjoys more advantages because it manages to exploit the global information to reconstruct all the frames in a video sequence.
  Overall, similar generator networks enjoy such a great advancement under the omniscient framework, which proves its effectiveness undoubtedly.

\subsection{Reconstruction Refinement} \label{sec:exp_refine}
  In Section \ref{sec:exp_frameworks}, because IVSR, RVSR and HVSR do not own a precursor network, we adopt the Bicubic amplified results to replace the SR frames given 
by the precursor for OVSR.
  However, the Bicubic amplification is not optimal for our omniscient framework.
  Since OVSR consists of a precursor network and a successor network, we naturally consider making the precursor and successor learn the structures and details of frames
respectively, and then add them for refinement.
  Thus, we change $\alpha$ in Equation \eqref{eq:loss} during training to explore the best option.

  Training curves are shown in Figure \ref{fig:abl_lovsr_p+s} and Figure \ref{fig:abl_govsr_p+s}, while the specific numbers are demonstrated Table \ref{table:abl_refine}.
  For LOVSR, we find $\alpha=0.1$ to be the optimal option, which surpasses the ``+Bicubic'' 0.02 dB.
  For GOVSR, setting $\alpha=0.01$ achieves 31.36 dB in PSNR, which surpasses the ``+Bicubic'' 0.09 dB.
  We reckon that $Net_p$ and $Net_s$ in GOVSR cooperate better due to the global information utilization.

  It is worth mentioning that in some case, especially when imposing no restrictions on the precursor ($\alpha=0$), as illustrated in Figure \ref{fig:abl_lovsr_p+s} 
and Figure \ref{fig:abl_govsr_p+s}, models could suffer sudden drops during training, but they will still converge eventually.

\subsection{Proportion of Precursor and Successor} \label{sec:exp_proportion}
  In Section \ref{sec:exp_frameworks} and Section \ref{sec:exp_refine}, we only set 1 PFRB in the precursor and 5 PFRBs in the successor for OVSR.
  Nevertheless, the optimal proportion of $Net_p$ and $Net_s$ still needs to be explored.
  Thus, we keep the total number of residual blocks fixed (6), and adjust the residual blocks in $Net_p$ (from 0 to 6) and $Net_s$ (from 6 to 0) to find 
an appropriate proportion.
  Training curves are shown in Figure \ref{fig:abl_lovsr_proportion} and Figure \ref{fig:abl_govsr_proportion}, while the specific numbers are demonstrated Table \ref{table:abl_proportion}.
  Obviously, neither ``0+6'' or ``6+0'' is the best or second-best option, which verifies the necessity of designing both the precursor and successor networks.
  Both LOVSR and GOVSR achieve their best performances with 4 residual blocks in the precursor and 2 in the successor, but GOVSR achieves 31.52 dB under the optimal setting, which again
proves the advantage of utilizing global information.
  Note that ``4:2'' is the optimal proportion under 6 residual blocks in total, however, there could be a better proportion if with more residual blocks.

  After optimization, our models LOVSR and GOVSR achieve 31.28 dB and 31.52 dB in PSNR, which surpass the HVSR 0.18 dB and 0.42 dB respectively.
  With similar parameters and calculation costs, our OVSR outperforms the existing IVSR, RVSR and HVSR enough to prove the effectiveness of further involving 
estimated hidden states from the present and future.

\begin{table*}[ht]
  \caption{PSNR (dB) / SSIM of different video SR models on Vid4 testing dataset \cite{VESPCN} by the upscaling factor of 4.
  \textcolor{red}{Red} and \textcolor{blue}{blue} respectively indicate the best and second-best results. The $^*$ denotes the results reported in the 
  original papers.}
  \renewcommand{\arraystretch}{1.2}
  \label{table:vid4}
  \centering
  \footnotesize
  \begin{tabular*}{0.98\textwidth}{@{\extracolsep{\fill}}ccccccc}
    \toprule
    Methods                  & calendar       & city           & foliage        & walk           & average        &average$^*$     \\
    \midrule
    DRVSR \cite{DRVSR}       & 22.88 / 0.7586 & 27.06 / 0.7698 & 25.58 / 0.7307 & 29.11 / 0.8876 & 26.16 / 0.7867 & 25.52 / 0.7600 \\
    FRVSR \cite{FRVSR}       & 23.46 / 0.7854 & 27.70 / 0.8099 & 25.96 / 0.7560 & 29.69 / 0.8990 & 26.70 / 0.8126 & 26.69 / 0.8220 \\
    MMCNN \cite{MMCNN}       & 23.63 / 0.7969 & 27.47 / 0.8083 & 26.01 / 0.7532 & 29.94 / 0.9030 & 26.76 / 0.8154 & 26.28 / 0.7844 \\
    MTUDM \cite{MTUDM}       & 23.76 / 0.8026 & 27.67 / 0.8145 & 26.08 / 0.7587 & 30.16 / 0.9069 & 26.92 / 0.8207 & 26.57 / 0.7989 \\
    DUF\_52L \cite{DUFVSR}   & 23.85 / 0.8052 & 27.97 / 0.8253 & 26.22 / 0.7646 & 30.47 / 0.9118 & 27.13 / 0.8267 & 27.34 / 0.8327 \\
    RBPN \cite{RBPN}         & 24.33 / 0.8244 & 28.28 / 0.8413 & 26.46 / 0.7753 & 30.58 / 0.9130 & 27.41 / 0.8385 & 27.16 / 0.8190 \\
    EDVR \cite{EDVR}         & 24.30 / 0.8242 & 28.04 / 0.8382 & 26.45 / 0.7744 & 30.63 / 0.9140 & 27.36 / 0.8377 & 27.35 / 0.8264 \\
    PFNL \cite{PFNL}         & 24.37 / 0.8246 & 28.09 / 0.8385 & 26.51 / 0.7768 & 30.64 / 0.9134 & 27.41 / 0.8383 & 27.40 / 0.8384 \\
    FFCVSR \cite{FFCVSR}     & 24.39 / 0.8250 & 27.80 / 0.8314 & 26.70 / 0.7868 & 30.55 / 0.9124 & 27.36 / 0.8389 & 26.97 / 0.8300 \\
    RLSP7-256 \cite{RLSP}    & 24.60 / 0.8335 & 28.14 / 0.8453 & 26.75 / 0.7925 & 30.88 / 0.9192 & 27.60 / 0.8476 & 27.55 / \hspace{8pt} - \hspace{7pt} \\
    TDAN \cite{TDAN}         & 23.56 / 0.7896 & 27.53 / 0.8028 & 26.00 / 0.7491 & 29.99 / 0.9032 & 26.77 / 0.8112 & 26.86 / 0.8140\\
    TGA$^*$ \cite{TGA}       & 24.47 / 0.8286 & 28.37 / 0.8419 & 26.59 / 0.7793 & 30.96 / 0.9181 & 27.59 / 0.8419 & 27.59 / 0.8419 \\
    RSDN9-128 \cite{RSDN}    & 24.74 / 0.8386 & 28.75 / 0.8554 & 27.00 / 0.8013 & 30.85 / 0.9183 & 27.83 / 0.8534 & 27.92 / 0.8505 \\
    \bottomrule
    LOVSR-4+2-56 (ours)      & 24.71 / 0.8378 & 28.47 / 0.8502 & 26.94 / 0.7969 & 30.97 / 0.9204 & 27.78 / 0.8513 & 27.78 / 0.8513 \\
    LOVSR-8+4-56 (ours)      & 24.93 / 0.8439 & \textcolor{blue}{29.08} / 0.8616 & 27.11 / 0.8073 & 31.23 / 0.9239 & 28.09 / 0.8592 & 28.09 / 0.8592 \\
    LOVSR-8+4-80 (ours)      & 25.10 / 0.8515 & 28.97 / 0.8666 & 27.25 / 0.8124 & 31.47 / 0.9273 & 28.20 / 0.8644 & 28.20 / 0.8644 \\
    GOVSR-4+2-56 (ours)      & 24.88 / 0.8463 & 28.74 / 0.8614 & 27.06 / 0.8075 & 31.27 / 0.9245 & 27.99 / 0.8599 & 27.99 / 0.8599 \\
    GOVSR-8+4-56 (ours)      & \textcolor{blue}{25.16} / \textcolor{blue}{0.8556} & 28.76 / \textcolor{blue}{0.8683} & \textcolor{blue}{27.36} / \textcolor{blue}{0.8190} & \textcolor{blue}{31.60} / \textcolor{blue}{0.9290} & \textcolor{blue}{28.22} / \textcolor{blue}{0.8680} & \textcolor{blue}{28.22} / \textcolor{blue}{0.8680}\\
    GOVSR-8+4-80 (ours)      & \textcolor{red}{25.28} / \textcolor{red}{0.8581} & \textcolor{red}{29.10} / \textcolor{red}{0.8769} & \textcolor{red}{27.49} / \textcolor{red}{0.8230} & \textcolor{red}{31.79} / \textcolor{red}{0.9314} & \textcolor{red}{28.41} / \textcolor{red}{0.8724} & \textcolor{red}{28.41} / \textcolor{red}{0.8724}\\
    \bottomrule
\end{tabular*}
\end{table*}

\begin{table*}[ht]
  \caption{PSNR (dB) / SSIM of different video SR models on UDM10 testing dataset \cite{MTUDM} by the upscaling factor of 4.
  \textcolor{red}{Red} and \textcolor{blue}{blue} respectively indicate the best and second-best results.}
  \renewcommand{\arraystretch}{0.8}
  \label{table:udm10}
  \centering
  \fontsize{6.3pt}{\baselineskip}\selectfont
  \setlength{\tabcolsep}{0.6mm}{
  \begin{tabular*}{0.98\textwidth}{@{\extracolsep{\fill}}cccccccccccc}
    \toprule
    Methods                  & archpeople   & archwall     & auditorium   & band         & caffe        & camera       & clap         & lake         & photography  & polyflow     & average      \\
    \midrule
    DRVSR \cite{DRVSR}       & 35.83/0.9547 & 41.16/0.9671 & 29.00/0.9039 & 34.32/0.9579 & 39.08/0.9715 & 45.19/0.9905 & 36.20/0.9635 & 31.15/0.8440 & 36.60/0.9627 & 37.91/0.9565 & 36.64/0.9472 \\
    FRVSR \cite{FRVSR}       & 36.24/0.9579 & 41.65/0.9710 & 29.81/0.9181 & 34.54/0.9589 & 39.82/0.9746 & 46.07/0.9912 & 36.51/0.9659 & 31.70/0.8623 & 36.95/0.9655 & 38.38/0.9597 & 37.17/0.9525 \\
    MMCNN \cite{MMCNN}       & 36.95/0.9636 & 42.12/0.9729 & 30.05/0.9217 & 35.23/0.9645 & 40.29/0.9760 & 46.89/0.9922 & 37.32/0.9704 & 31.76/0.8642 & 37.81/0.9704 & 38.85/0.9649 & 37.73/0.9561 \\
    MTUDM \cite{MTUDM}       & 37.16/0.9655 & 42.33/0.9744 & 30.37/0.9274 & 35.46/0.9661 & 40.68/0.9773 & 47.15/0.9924 & 37.69/0.9727 & 32.03/0.8734 & 38.18/0.9727 & 39.10/0.9670 & 38.02/0.9589 \\
    DUF\_52L \cite{DUFVSR}   & 36.92/0.9638 & 42.53/0.9754 & 30.27/0.9257 & 35.49/0.9660 & 41.03/0.9785 & 47.30/0.9927 & 37.70/0.9719 & 32.06/0.8730 & 38.02/0.9719 & 39.25/0.9667 & 38.05/0.9586 \\
    RBPN \cite{RBPN}         & 38.50/0.9729 & 43.53/0.9790 & 31.23/0.9376 & 35.49/0.9678 & 41.83/0.9810 & 49.25/0.9940 & 38.35/0.9757 & 32.48/0.8837 & 38.96/0.9771 & 40.38/0.9732 & 39.00/0.9642 \\
    EDVR \cite{EDVR}         & 38.46/0.9732 & 43.35/0.9783 & 31.15/0.9372 & 35.97/0.9696 & 41.76/0.9808 & 49.49/0.9947 & 38.22/0.9759 & 32.21/0.8790 & 39.40/0.9793 & 40.47/0.9739 & 39.05/0.9642 \\
    PFNL \cite{PFNL}         & 38.35/0.9724 & 43.55/0.9792 & 31.18/0.9369 & 36.01/0.9691 & 41.84/0.9808 & 49.26/0.9941 & 38.33/0.9756 & 32.53/0.8865 & 38.95/0.9768 & 40.04/0.9734 & 39.00/0.9645 \\
    FFCVSR \cite{FFCVSR}     & 37.50/0.9667 & 42.98/0.9766 & 30.50/0.9270 & 35.71/0.9669 & 41.27/0.9798 & 48.65/0.9936 & 37.88/0.9727 & 32.23/0.8729 & 38.42/0.9739 & 39.74/0.9691 & 38.49/0.9599 \\
    RLSP7-256 \cite{RLSP}    & 38.05/0.9704 & 43.46/0.9787 & 31.01/0.9342 & 36.05/0.9693 & 42.06/0.9818 & 49.14/0.9939 & 38.41/0.9756 & 32.60/0.8865 & 39.03/0.9771 & 40.38/0.9748 & 39.02/0.9642 \\
    TDAN \cite{TDAN}         & 37.95/0.9699 & 42.60/0.9747 & 30.54/0.9283 & 35.23/0.9645 & 40.59/0.9773 & 48.38/0.9936 & 37.42/0.9714 & 31.87/0.8668 & 38.28/0.9740 & 39.00/0.9660 & 38.19/0.9586 \\
    RSDN9-128 \cite{RSDN}    & 38.36/0.9719 & 43.68/0.9796 & 31.65/0.9416 & 36.13/0.9696 & 42.22/0.9824 & 49.88/0.9946 & 38.48/0.9762 & 32.67/0.8860 & 39.47/0.9793 & 40.44/0.9735 & 39.30/0.9655 \\
    \bottomrule
    LOVSR-4+2-56 (ours)      & 38.26/0.9718 & 43.57/0.9794 & 31.22/0.9376 & 36.23/0.9702 & 42.28/0.9825 & 49.43/0.9942 & 38.51/0.9763 & 32.73/0.8905 & 39.24/0.9781 & 40.43/0.9753 & 39.19/0.9656 \\
    LOVSR-8+4-56 (ours)      & 38.61/0.9736 & 43.84/0.9804 & 31.78/0.9437 & 36.52/0.9717 & 42.79/0.9836 & 50.28/0.9948 & 38.95/0.9781 & 32.89/0.8932 & 39.89/0.9807 & 40.99/0.9773 & 39.65/0.9677 \\
    LOVSR-8+4-80 (ours)      & \textcolor{red}{39.01}/\textcolor{blue}{0.9755} & \textcolor{blue}{44.13}/0.9815 & 32.24/0.9480 & 36.80/0.9732 & \textcolor{red}{43.16}/\textcolor{blue}{0.9843} & \textcolor{blue}{50.52}/\textcolor{red}{0.9950} & 39.26/0.9793 & 33.13/0.9006 & \textcolor{blue}{40.29}/0.9820 & \textcolor{red}{41.32}/\textcolor{blue}{0.9785} & \textcolor{blue}{39.99}/0.9698 \\
    GOVSR-4+2-56 (ours)      & 38.44/0.9727 & 43.73/0.9802 & 31.48/0.9408 & 36.32/0.9708 & 42.58/0.9833 & 49.54/0.9942 & 38.80/0.9781 & 32.89/0.8969 & 39.44/0.9790 & 40.45/0.9767 & 39.37/0.9673 \\
    GOVSR-8+4-56 (ours)      & \textcolor{blue}{38.79}/0.9747 & 44.12/\textcolor{blue}{0.9818} & \textcolor{blue}{32.30}/\textcolor{blue}{0.9488} & \textcolor{blue}{36.82}/\textcolor{blue}{0.9734} & \textcolor{blue}{43.09}/0.9841 & 50.46/\textcolor{blue}{0.9949} & \textcolor{blue}{39.35}/\textcolor{blue}{0.9802} & \textcolor{blue}{33.22}/\textcolor{blue}{0.9044} & \textcolor{blue}{40.29}/\textcolor{blue}{0.9821} & 41.17/\textcolor{red}{0.9791} & 39.96/\textcolor{blue}{0.9703} \\
    GOVSR-8+4-80 (ours)      & \textcolor{red}{39.01}/\textcolor{red}{0.9757} & \textcolor{red}{44.44}/\textcolor{red}{0.9828} & \textcolor{red}{32.53}/\textcolor{red}{0.9510} & \textcolor{red}{37.03}/\textcolor{red}{0.9744} & 43.07/\textcolor{red}{0.9845} & \textcolor{red}{50.60}/\textcolor{red}{0.9950} & \textcolor{red}{39.58}/\textcolor{red}{0.9809} & \textcolor{red}{33.34}/\textcolor{red}{0.9066} & \textcolor{red}{40.48}/\textcolor{red}{0.9830} & \textcolor{blue}{41.28}/\textcolor{red}{0.9791} & \textcolor{red}{40.14}/\textcolor{red}{0.9713} \\
    \bottomrule
  \end{tabular*}}
\end{table*}

\begin{table*}[ht]
  \caption{Comprehensive comparisons of different video SR methods.}
  \renewcommand{\arraystretch}{1.2}
  \label{table:comparison_sota}
  \centering
  \footnotesize
  \begin{tabular*}{0.98\textwidth}{@{\extracolsep{\fill}}cccccccc}
    \toprule
    Methods                  & Framework  & Generator  & Frame & Parameter (M)           & \tabincell{c}{FLOPs (T) \\ $1280\times 720$} & \tabincell{c}{Time (ms) / FPS\\ $1280\times 720$}              & \tabincell{c}{Time (ms) / FPS\\ $1920\times 1080$}                  \\
    \midrule
    DRVSR \cite{DRVSR}       & iterative  & recurrent  & 3     & \textcolor{red}{1.722}  & 0.415                                        & \hspace{6pt} 123.6  / \hspace{2pt} 8.09                        & \hspace{6pt} 276.2   / \hspace{2pt} 3.62                            \\
    FRVSR \cite{FRVSR}       & recurrent  & recurrent  & 2     & 5.058                   & 0.348                                        & \hspace{10pt} 77.5  / 12.90                                    & \hspace{6pt} 172.1   / \hspace{2pt} 5.81                            \\
    MMCNN \cite{MMCNN}       & iterative  & recurrent  & 5     & 10.582 ~                & 3.347                                        & \hspace{6pt} 696.7  / \hspace{2pt} 1.44                        & \hspace{2pt} 1548.9  / \hspace{2pt} 0.65                            \\
    MTUDM \cite{MTUDM}       & iterative  & recurrent  & 5     & 5.919                   & 1.672                                        & \hspace{6pt} 506.9  / \hspace{2pt} 1.97                        & \hspace{2pt} 1127.6  / \hspace{2pt} 0.89                            \\
    DUF\_52L \cite{DUFVSR}   & iterative  & iterative  & 7     & 5.824                   & 2.348                                        & \hspace{2pt} 1108.0 / \hspace{2pt} 0.90                        & \hspace{2pt} 2499.8  / \hspace{2pt} 0.40                            \\
    RBPN \cite{RBPN}         & iterative  & recurrent  & 7     & 12.772 ~                & 8.516                                        & \hspace{2pt} 6555.7 / \hspace{2pt} 0.15                        &              14935.8 / \hspace{2pt} 0.07                            \\
    EDVR \cite{EDVR}         & iterative  & iterative  & 7     & 20.699 ~                & 2.954                                        & \hspace{6pt} 436.7  / \hspace{2pt} 2.29                        & \hspace{6pt} 979.2   / \hspace{2pt} 1.02                            \\
    PFNL \cite{PFNL}         & iterative  & iterative  & 7     & 3.003                   & 0.940                                        & \hspace{6pt} 231.0  / \hspace{2pt} 4.33                        & \hspace{6pt} 567.0   / \hspace{2pt} 1.76                            \\
    FFCVSR \cite{FFCVSR}     & hybrid     & hybrid     & 3     & 5.581                   & 0.322                                        & \hspace{10pt} 50.8  / 19.69                                    & \hspace{6pt} 112.7   / \hspace{2pt} 8.87                            \\
    RLSP7-256 \cite{RLSP}    & hybrid     & hybrid     & 3     & 5.553                   & 0.320                                        & \hspace{10pt} \textcolor{blue}{42.9}  / \textcolor{blue}{23.31}  & \hspace{10pt} \textcolor{blue}{91.9}   / \textcolor{blue}{10.88}  \\
    TDAN \cite{TDAN}         & iterative  & iterative  & 5     & 2.285                   & 0.558                                        & \hspace{6pt} 102.3  / \hspace{2pt} 9.78                        & \hspace{6pt} 225.4   / \hspace{2pt} 4.44                            \\
    TGA \cite{TGA}           & iterative  & iterative  & 7     & 7.058                   & 0.700                                        & \hspace{6pt} 383.5  / \hspace{2pt} 2.61                        & \hspace{6pt} 869.7   / \hspace{2pt} 1.15                     \\
    RSDN9-128 \cite{RSDN}    & recurrent  & recurrent  & 2     & 6.180                   & 0.356                                        & \hspace{10pt} 59.3  / 16.86                                    & \hspace{6pt} 132.0   / \hspace{2pt} 7.58                            \\
    \bottomrule
    OVSR-4+2-56 (ours)       & omniscient & omniscient & 3     & \textcolor{blue}{1.897} & \textcolor{red}{0.110}                       & \hspace{10pt} \textcolor{red}{25.4} / \textcolor{red}{39.37}   & \hspace{10pt} \textcolor{red}{56.1}   / \textcolor{red}{17.83}      \\
    OVSR-8+4-56 (ours)       & omniscient & omniscient & 3     & 3.480                   & \textcolor{blue}{0.201}                      & \hspace{10pt} 46.5  / 21.51                                    & \hspace{6pt} 101.9   / \hspace{2pt} 9.81                            \\
    OVSR-8+4-80 (ours)       & omniscient & omniscient & 3     & 7.062                   & 0.407                                        & \hspace{10pt} 81.2  / 12.32                                    & \hspace{6pt} 178.5   / \hspace{2pt} 5.60                            \\
    \bottomrule
\end{tabular*}
\end{table*}

\begin{figure*}[ht]
  \centering
  \includegraphics[width=0.85\textwidth]{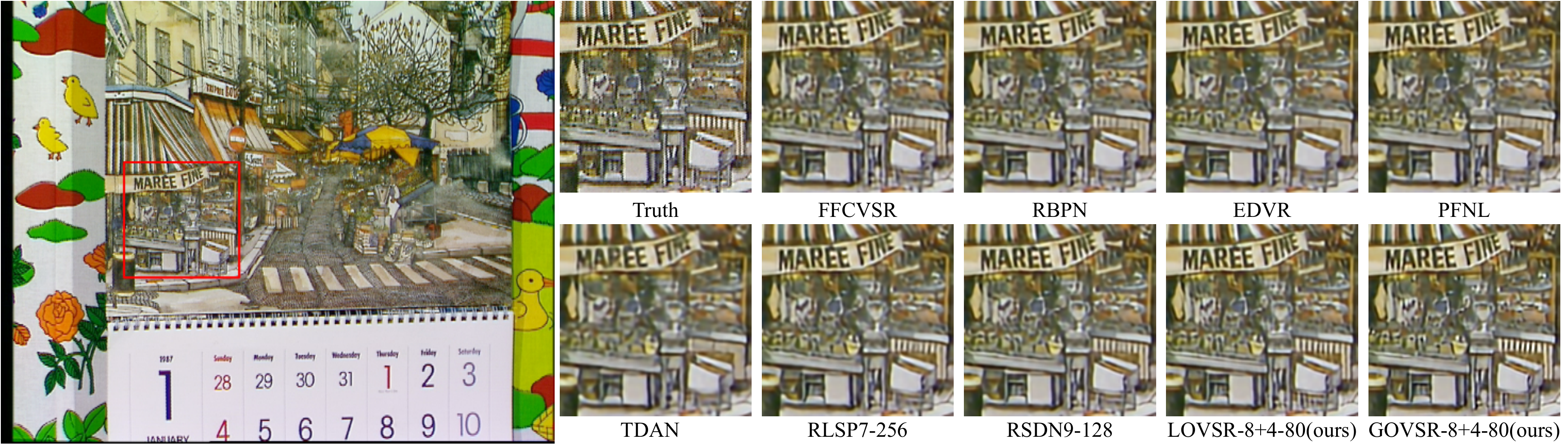}
  \caption{Results of different methods on \emph{calendar} from Vid4 \cite{VESPCN} dataset.}
  \label{fig:compare}
\end{figure*}

\subsection{Comparisons with SOTA Methods} \label{sec:exp_comp}
  Because most methods train their models on different training datasets with different down-sampling kernels, it is unfair to compare these methods 
only according to their papers.
  Thus, we have reimplemented dozens of SOTA VSR methods on the same training dataset with the same down-sampling kernels.
  Since PFNL \cite{PFNL} has already rebuilt DRVSR \cite{DRVSR}, FRVSR \cite{FRVSR} and DUF\_52L \cite{DUFVSR} under the same training conditions, we adopt 
their public codes.
  Based on the public code of PFNL, we reimplement MMCNN \cite{MMCNN} and MTUDM \cite{MTUDM} on the TensorFlow platform.
  Moreover, We rebuild RBPN \cite{RBPN}, EDVR \cite{EDVR}, FFCVSR \cite{FFCVSR}, TDAN \cite{TDAN}, RSDN \cite{RSDN} and RLSP \cite{RLSP} on the PyTorch platform. 
  We could not retrain TGA because it requires too much GPU memory, and we have to report the results in its paper.
  
  We first reimplement these VSR methods on MM522 \cite{MMCNN} training dataset and then test them on Vid4 \cite{VESPCN} and UDM10 \cite{MTUDM} testing 
dataset.
  The PSNR and SSIM values are calculated only on the luminance channel of YCbCr colorspace, skipping the first and last two frames and eliminating 8 
pixels on four borders of each frame \cite{VSRnet, PFNL}.
  As illustrated in Table \ref{table:vid4}, our heavy model GOVSR-8+4-80 has achieved the best result, while our medium model GOVSR-8+4-56 achieves the 
second-best performance.
  LOVSR-based models behave a little worse than GOVSR-based models because they can not utilize the global information, but they still outperform 
most other SOTA methods.

  Because Vid4 dataset contains only 4 scenes at a low resolution, we further test these methods on a bigger testing dataset UDM10.
  As many methods have not conducted experiments on UDM10 dataset, we only report the results rebuilt by us in Table \ref{table:udm10}.
  Our heavy model GOVSR-8+4-80 achieves the best performance, while GOVSR-8+4-56 and LOVSR-8+4-80 achieve comparable performances.
  To make a comprehensive comparison, we show more details of these methods, \eg type of framework and generator, frame number, parameters, FLOPs, testing 
time cost and frames per second (FPS).
  As demonstrated in Table \ref{table:comparison_sota}, our light model OVSR-4+2-56 achieves the fastest speed and capable of real-time $4\times$ VSR for 720p.
  Our medium model OVSR-8+4-56 surpasses all of the other SOTA methods a lot in PSNR and is still faster than almost all of them.
  Our heavy model OVSR-8+4-80 has to sacrifice the speed for achieving the best performance, which is necessary and worthwhile.
  Also shown in Figure \ref{fig:testtime}, the iterative methods suffer from huge calculation and time costs, although the recurrent and hybrid methods 
are faster, they are still limited in performance.
  As illustrated in Figure \ref{fig:compare}, our methods can restore clearer details.
  In all, our methods are superior to these SOTA VSR counterparts in objective metrics, subjective visual effects and complexity.
  Please refer to the appendix for more ablation studies and comparisons.

\section{Conclusion} \label{sec:conclusion}
  In this paper, we propose an omniscient framework for VSR, which can be divided into the local omniscient VSR and global omniscient VSR.
  LOVSR suits online VSR and GOVSR fits offline VSR, and they both leverage LR frames and estimated hidden states from the past, present and future.
  We have conducted abundant experiments to prove the robustness of the omniscient framework, which is superior to other frameworks under similar complexity.
  Combined with a sophisticated generator network, the omniscient framework makes it possible to conduct VSR in real-time but still with an amazing 
performance.
  Due to the overwhelming advantages of this framework, we hope it will become a standard framework in VSR, based on which the researchers are welcome to 
design more effective generators, motion information capturing modules, as well as loss functions to further unleash its potential.


{\small
\bibliographystyle{ieee_fullname}
\bibliography{egbib}
}

\clearpage

\begin{appendices}
  \noindent \textbf{\Large Appendix}

  We conduct more ablation stuides in Section \ref{sec:app_abl}. 
  We show the result on Vimeo-90K \cite{TOFLOW} in Section \ref{sec:app_vimeo} and more visual comparisons in Section \ref{sec:app_visual}.
  \section{Ablation Studies} \label{sec:app_abl}

  \subsection{HVSR with 4 LR frames}
\begin{figure}[ht]
      \centering
      \includegraphics[width=0.4\textwidth]{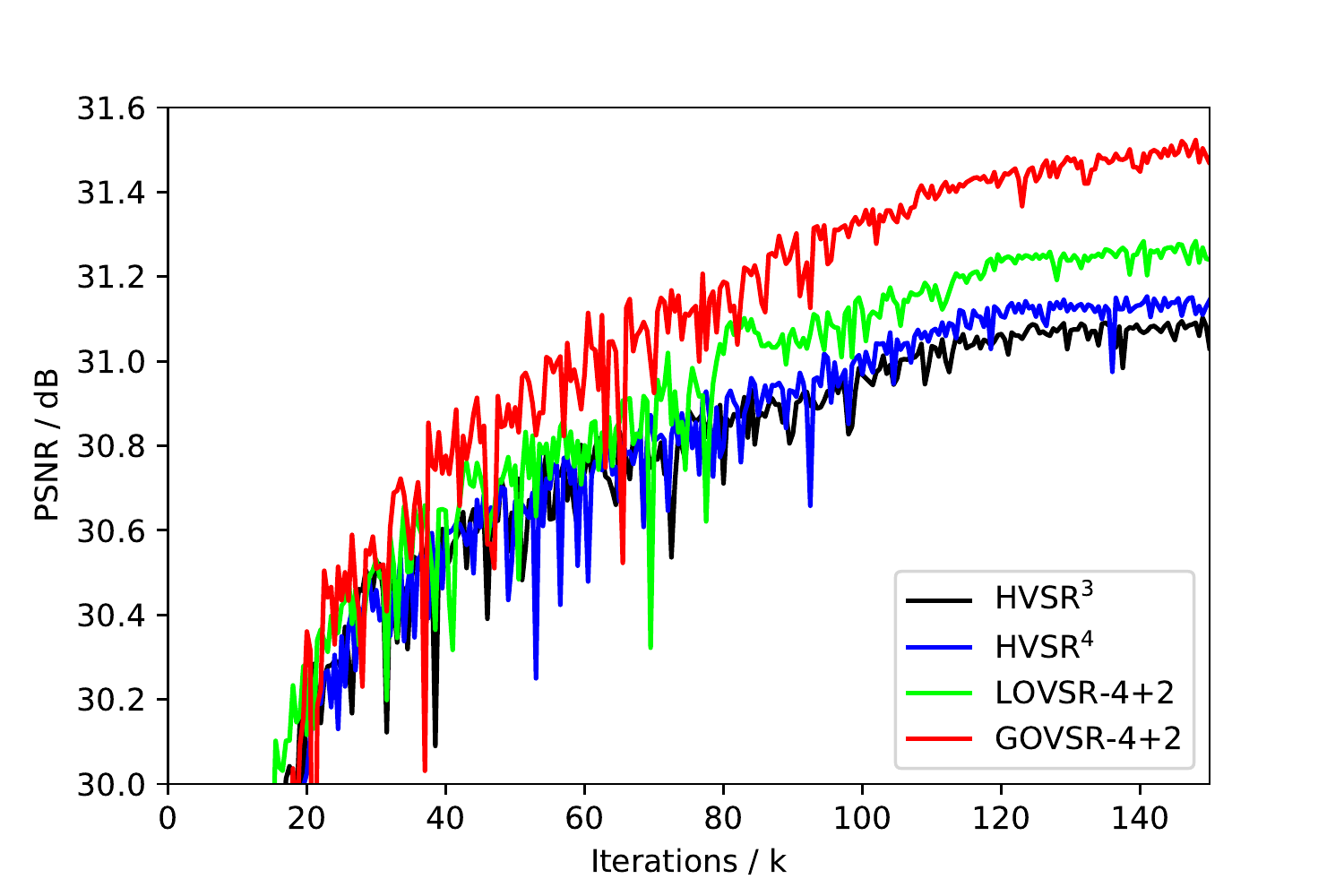}
      \caption{Training curves of different models.}
      \label{fig:app_ablation}
  \end{figure}

    Our LOVSR actually utilizes 4 LR frames through the future hidden states $H_{t+1}$, and thus we have conducted an additional experiment of HVSR adopting 4 LR frames ($I_{t-1}, I_t, I_{t+1}, I_{t+2}$).
    As shown in Figure \ref{fig:app_ablation} and Table \ref{table:app_ablation}, HVSR$^3$ and HVSR$^4$ denote HVSR with 3 and 4 input LR frames respectively, where HVSR$^4$ surpasses HVSR$^3$ about 0.05 dB.
    Still, with the same 4 LR frames, our LOVSR outperforms HVSR$^4$ about 0.13 dB, and our GOVSR surpasses HVSR$^4$ about 0.37 dB with the assist of all LR frames in a video sequence.
    Note that these models have a similar number of parameters and calculation costs, which again proves the robustness of our OVSR framework.

    \begin{table}[ht]
    \caption{PSNR (dB) of different models.}
    \renewcommand{\arraystretch}{1.2}
    \label{table:app_ablation}
    \centering
    \footnotesize
    \begin{tabular}{@{\extracolsep{\fill}}ccccccccc}
      \toprule
      Model              & HVSR$^3$      & HVSR$^4$      & LOVSR-4+2     & GOVSR-4+2 \\
      \midrule
      PSNR (dB)          & 31.10         & 31.15         & 31.28         & 31.52\\
      \bottomrule
    \end{tabular}
    \end{table}

  \subsection{Influences of Input Information}
    We further investigate the influence of each input information on the performance, as demonstrated in Table \ref{table:app_abl_framework}, we present the 
  performances of different models by removing part of the input information.

    Obviously, all models drop seriously without the center LR frame $I_t$, which accords with common sense because $I_t$ contains the most basic
  and significant source information.
    Interestingly, IVSR, RVSR, and HVSR models all achieve 26.61 dB without $I_t$, and this is almost the same as the Bicubic magnified center frame $I_t^{Bic}$ actually.
    Because generating the Bicubic magnified center frame $I_t^{Bic}$ does not require learnable parameters, the IVSR, RVSR, and HVSR models all drop to $I_t^{Bic}$ without $I_t$.
    Nevertheless, our OVSR models add the SR frames by $Net_p$ and $Net_s$ for reconstruction refinement (discussed in Equation (3) and Section 4.3 in the original paper), which requires 
  the cooperation of two learnable networks and thus is more unstable than the Bicubic magnification, where they basically drop more than 10 dB by removing the center frame $I_t$.

    IVSR drops nearly the same by removing $I_{t-1}$ or $I_{t+1}$, which means $I_{t-1}$ and $I_{t+1}$ contribute almost equally to the result, and thus 
  it is unwise for RVSR to overlook the subsequent frame $I_{t+1}$.

    Excluding $I_{t-1}$ or $H_{t-1}$, RVSR drops a lot, which proves that the hidden states can provide some beneficial information.

    Moreover, HVSR drops more by removing $I_{t+1}$ compared to removing $I_{t-1}$, and we reckon that this is due to the assist from $H_{t-1}$.
    This phenomenon further confirms that the hidden states contribute to the VSR indeed, and thus it makes sense to further adopt the hidden states from the 
  present and future to help VSR.
    Similar phenomena can also be observed in our models LOVSR and GOVSR.

    Amazingly, by removing $I_{t-1}$ in $Net_s$, our GOVSR stays the same but LOVSR increases 0.01 dB in PSNR instead, and we owe it to the OVSR framework, which leverages hidden 
  states from the past, present and future to help VSR, through which it basically does not need $I_{t-1}$ in $Net_s$ anymore.
    Last but not least, compared to removing the LR frames $I$, our models LOVSR and GOVSR deteriorate more seriously or comparably by removing the corresponding hidden states $H$ in $Net_s$, which 
  again confirms that our models indeed make good use of the hidden states from the past, present and future.

  \begin{table*}[ht]
    \caption{PSNR (dB) of different models, where `Full' denotes the original model with all input information, and `w/o' means without.}
    \renewcommand{\arraystretch}{1.2}
    \label{table:app_abl_framework}
    \centering
    \scriptsize
    \begin{tabular}{@{\extracolsep{\fill}}cccc|ccc|ccc}
      \toprule
      Model                      & IVSR           & RVSR           & HVSR            & \multicolumn{3}{c|}{LOVSR-4+2}                  & \multicolumn{3}{c}{GOVSR-4+2}                     \\
      Network                    & G              & G              & G               & $Net_p$        &$Net_s$         & Both          & $Net_p$         &$Net_s$         & Both           \\
      \midrule
      Full                       & 30.66          & 30.60          & 31.10           & 31.28          & 31.28          & 31.28         & 31.52           & 31.52          & 31.52          \\
      w/o $I_{t-1}$ \hspace{0pt} & 29.07 (-1.59)  & 28.95 (-1.65)  & 29.18 (-1.92)   & 29.27 (-2.01)  & 31.29 (+0.01)  & 29.27 (-2.01) & 28.92 (-2.60)   & 31.52 (-0.00)  & 28.91 (-2.61)  \\
      w/o $I_{t}$ \hspace{9pt}   & 26.61 (-4.05)  & 26.61 (-3.99)  & 26.61 (-4.49)   & 21.18 (-10.10) & 16.78 (-14.50) & 8.75 (-22.53) & 22.16 (-9.36)   & 13.70 (-17.82) & 8.75  (-22.77) \\
      w/o $I_{t+1}$ \hspace{1pt} & 29.10 (-1.56)  & -              & 28.91 (-2.19)   & 28.94 (-2.34)  & 30.66 (-0.62)  & 28.60 (-2.68) & 28.98 (-2.54)   & 30.90 (-0.62)  & 28.87 (-2.65)  \\
      w/o $H_{t-1}$              & -              & 29.35 (-1.25)  & 29.81 (-1.29)   & 29.74 (-1.54)  & 31.01 (-0.27)  & 29.53 (-1.75) & 29.59 (-1.93)   & 30.85 (-0.67)  & 29.39 (-2.13)  \\
      w/o $H_{t}$ \hspace{7pt}   & -              & -              & -               & -              & 11.18 (-20.10) & -             & -               & 12.44 (-19.08) & -              \\
      w/o $H_{t+1}$              & -              & -              & -               & -              & 30.52 (-0.76)  & -             & -               & 30.99 (-0.53)  & -              \\
      \bottomrule
      
    \end{tabular}
  \end{table*}

  \begin{table*}[ht]
    \caption{PSNR (dB) / SSIM of different video SR methods on Vimeo-90K testing dataset \cite{TOFLOW} by the upscaling factor of 4.
    \textcolor{red}{Red} and \textcolor{blue}{blue} respectively indicate the best and second-best results. The $^*$ denotes the results reported in the 
    original papers.}
    \renewcommand{\arraystretch}{1.2}
    \label{ttable:app_vimeo90k}
    \centering
    \footnotesize
    \begin{tabular*}{0.98\textwidth}{@{\extracolsep{\fill}}ccccccc}
      \toprule
      Methods                  & Vimeo-Slow     & Vimeo-Medium   & Vimeo-Fast     & Vimeo-All        \\
      \midrule
      RBPN$^*$ \cite{RBPN}     &34.18 / 0.9200  & 37.28 / 0.9470 & 40.03 / 0.9600 & 37.07 / 0.9435  \\
      EDVR$^*$ \cite{EDVR}     &-               &-               &-               & \textcolor{blue}{37.61} / 0.9489   \\
      FFCVSR \cite{FFCVSR}     & 33.59 / 0.9130 & 36.51 / 0.9416 & 38.68 / 0.9481 & 36.24 / 0.9367 \\
      RLSP7-256 \cite{RLSP}    & 33.86 / 0.9173 & 36.97 / 0.9463 & 39.20 / 0.9535 & 36.67 / 0.9415 \\
      RSDN9-128 \cite{RSDN}    & 33.91 / 0.9179 & 37.03 / 0.9466 & 39.41 / 0.9555 & 36.76 / 0.9421 \\
      \bottomrule
      LOVSR-8+4-80 (ours)      & \textcolor{blue}{34.51} / \textcolor{blue}{0.9256} & \textcolor{blue}{37.84} / \textcolor{blue}{0.9538} & \textcolor{blue}{40.23} / \textcolor{blue}{0.9615} & 37.53 / \textcolor{blue}{0.9492} \\
      GOVSR-8+4-80 (ours)      & \textcolor{red}{34.60} / \textcolor{red}{0.9270} & \textcolor{red}{37.95} / \textcolor{red}{0.9548} & \textcolor{red}{40.32} / \textcolor{red}{0.9624} & \textcolor{red}{37.63} / \textcolor{red}{0.9503} \\
      \bottomrule
  \end{tabular*}
  \end{table*}

  \begin{figure*}[!ht]
    \centering
    \begin{subfigure}[t]{0.95\textwidth}
      \centering
      \includegraphics[width=1\textwidth]{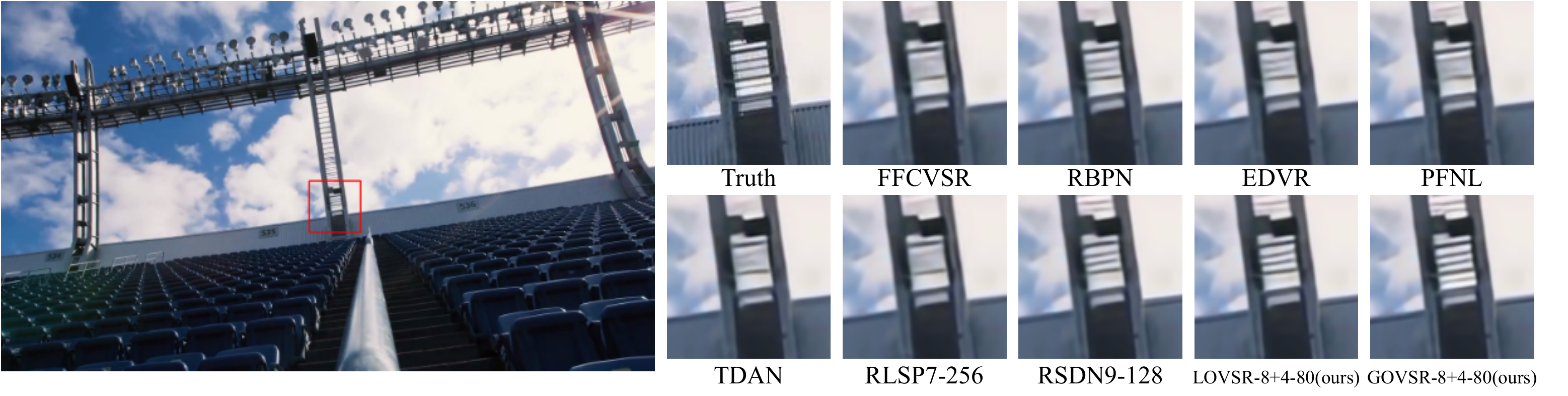}
      \caption{Frame 022 of auditorium from UDM10.}\label{fig:auditorium}
    \end{subfigure}

    \begin{subfigure}[t]{0.95\textwidth}
        \centering
        \includegraphics[width=1\textwidth]{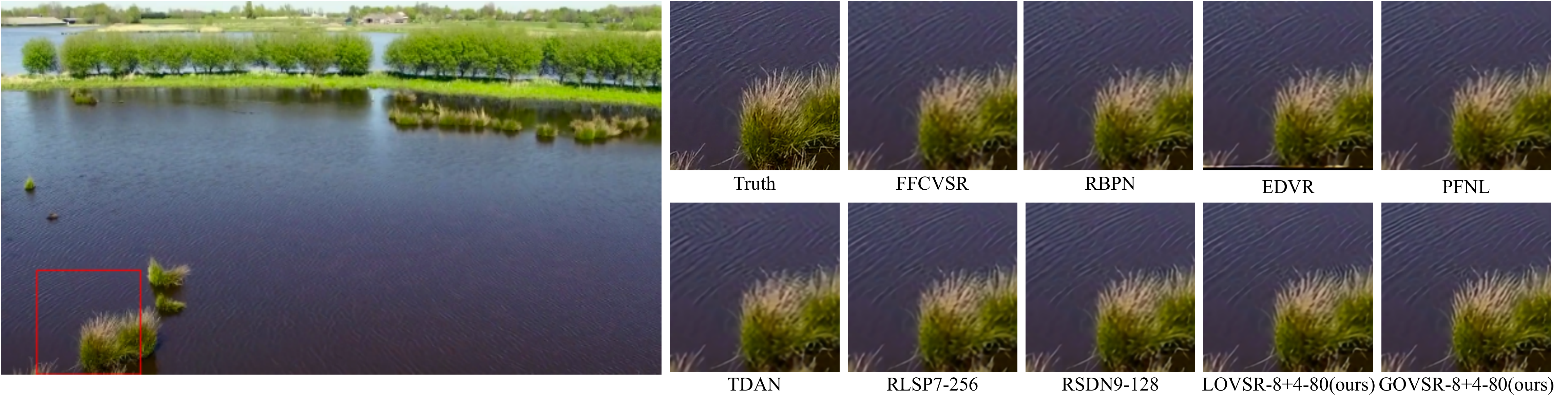}
        \caption{Frame 015 of lake from UDM10.}\label{fig:lake}
    \end{subfigure}

    \caption{Visual comparisons of different methods.}
    \label{fig:app_compare}
  \end{figure*}

    \begin{figure*}[!ht]
    \centering

      \begin{subfigure}[t]{0.95\textwidth}
          \centering
          \includegraphics[width=1\textwidth]{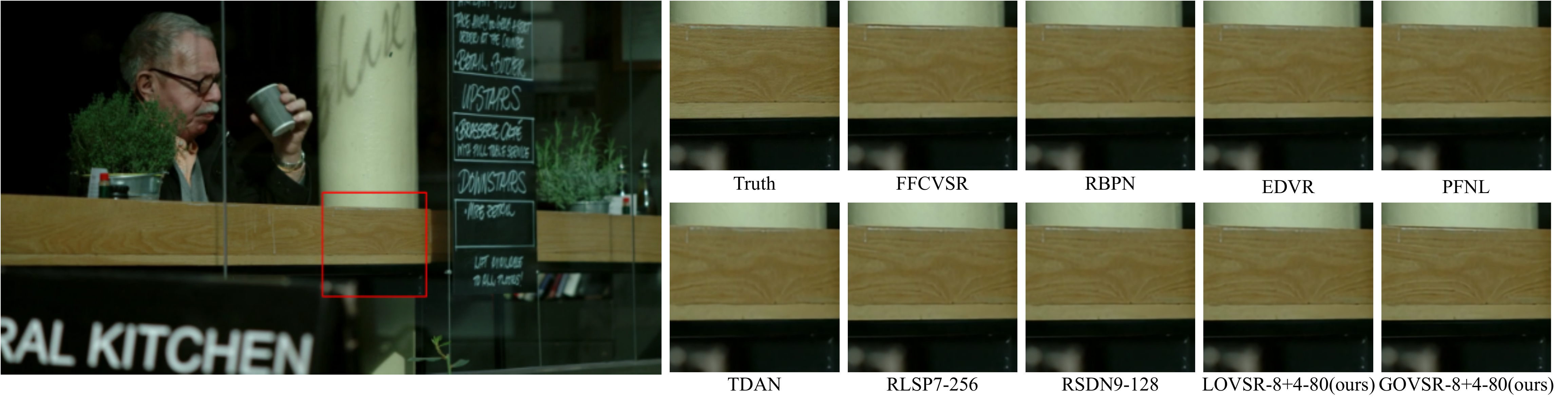}
          \caption{Frame 015 of caffe from UDM10.}\label{fig:caffe}
      \end{subfigure}

      \begin{subfigure}[t]{0.95\textwidth}
        \centering
        \includegraphics[width=1\textwidth]{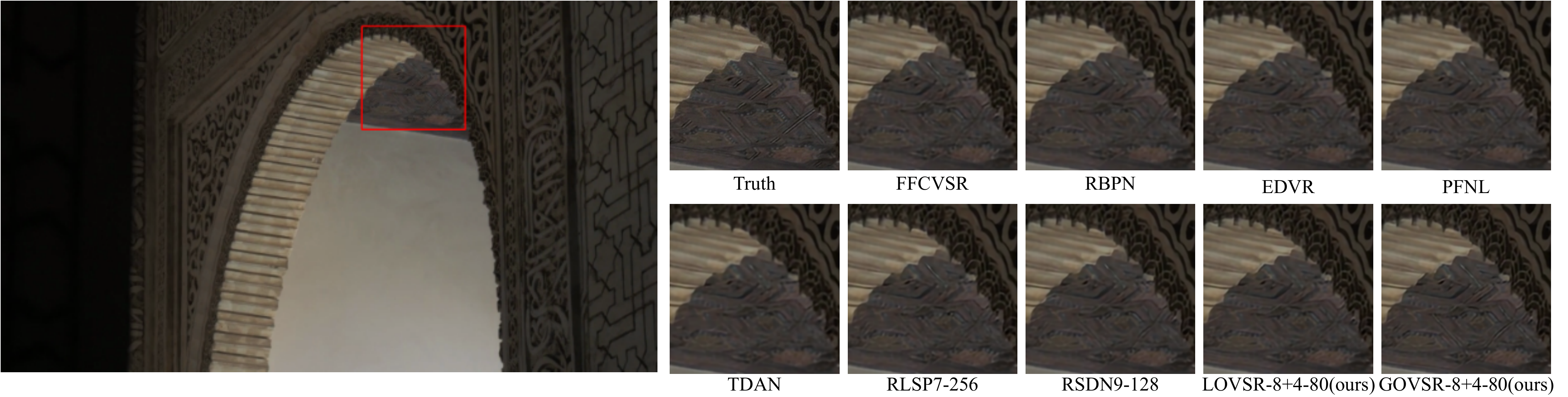}
        \caption{Frame 015 of archwall from UDM10.}\label{fig:archwall}
      \end{subfigure}

      \caption{Visual comparisons of different methods.}
      \label{fig:app_compare2}
    \end{figure*}

  \section{Result on Vimeo-90K Dataset} \label{sec:app_vimeo}
    We further conduct experiments on another public training dataset Vimeo-90K \cite{TOFLOW}, and test the models on its testing dataset Vimeo-90K-T.
    According to the average motion flow magnitude, Vimeo-90K-T is divided into 3 categories: slow, medium, and fast \cite{RBPN}, where there 
  are 1616, 4983, and 1225 sequences in each category.
    Albeit Vimeo-90K contains tens of thousands of video sequences, each of which consists of only 7 frames, and the HR frames are at the fixed 
  resolution $448\times 256$, which is quite low.
    Besides, during testing, other methods only calculate the PSNR/SSIM of the center SR frame, and consequently, we do not think this dataset is suitable 
  for VSR.
    Still, we retrain as many methods as we can on Vimeo-90K in a limited time for a more comprehensive comparison, under the same training settings.
    The PSNR and SSIM values are calculated only on the luminance channel of YCbCr colorspace, focusing on the center frame and eliminating 8 
  pixels on four borders.

    As shown in Table \ref{ttable:app_vimeo90k}, our model GOVSR-8+4-80 still achieves the best performance.
    It is worth mentioning that compared to GOVSR-8+4-80, LOVSR-8+4-80 can not utilize all 7 frames to rebuild the center frame, and thus it behaves a little worse.

  \section{Visual Comparisons} \label{sec:app_visual}
    We show more visual comparisons, as illustrated in Figure \ref{fig:app_compare} and Figure \ref{fig:app_compare2}, most methods can only recover the low-frequency contour of the 
  objects, which seems smooth and blurry.
    Our models can recover the right textures with more realistic details.
\end{appendices}

\end{document}